\documentclass[a4paper,11pt]{article}
\usepackage{jheppub} 
\usepackage{lineno}
\usepackage{tikz}
\usepackage{caption}
\usepackage{subcaption}
\usepackage{cancel}
\usepackage{csquotes}
\usepackage{layout}
\usepackage{mathtools}
\usepackage{array}
\newcolumntype{P}[1]{>{\centering\arraybackslash}p{#1}}

\renewcommand\slash[1]{\cancel{#1}}
\newcommand{\MSbar}{\overline{\text{MS}}}
\newcommand\order[1]{\mathcal{O}\left({#1}\right)}
\graphicspath{{./Figures/}}

\title{\boldmath Slepton pair production at next-to-leading power}

\affiliation[a]{Department of Physics, University of Oslo, 0316 Oslo, Norway}

\author[a]{Lasse Lorentz Braseth,}
\author[a]{Tore Klungland,}
\author[a]{and Are Raklev}

\emailAdd{lasselb@fys.uio.no}
\emailAdd{torekl@fys.uio.no}
\emailAdd{ahye@fys.uio.no}

\abstract{Near threshold, cross sections for the production of heavy particles are sensitive to large logarithmic terms, which must be resummed to all orders in perturbation theory. Current state-of-the art calculations for inclusive slepton pair production at hadron colliders has focused on higher-order logarithms in the leading power of the threshold variable. Here, we evaluate the next-to-leading power contribution in the threshold variable to leading logarithmic accuracy. We find that the next-to-leading power contributions can be significant compared to the next-to-leading logarithmic terms at leading power, and that existing calculations underestimate the scale error for large slepton masses. We include results for a potential future FCC-hh machine at $\sqrt{s}=85$ TeV.}

\begin{document}
\maketitle
\flushbottom

\section{Introduction}
\label{sec:intro}
As experimental data from the Large Hadron Collider (LHC) becomes increasingly precise and continues to show no clear evidence of new physics, the demand for more refined theoretical predictions both within and beyond the Standard Model (SM) grows stronger. Beyond higher fixed order terms, the effects of resummed contributions to hadronic cross sections are significant and crucial for precision comparison between theory and experiment in the SM. This has proven useful to capture effects at the production threshold~\cite{Altarelli:1979ub,Parisi:1979xd,Catani:1989ne,Sterman:1986aj}, for low transverse momentum~\cite{Dokshitzer:1978yd,Parisi:1979se,Collins:1984kg}, and jointly both of these effects~\cite{Li:1998is,Laenen:2000ij}. 

Today, state-of-the-art calculations in the SM has reached an impressive next-to-next-to-next-to-leading-logarithmic (N$^3$LL) accuracy at so-called leading power for threshold resummation~\cite{Moch:2005ky,Becher:2007ty}, and N$^4$LL for transverse-momentum resummation~\cite{Camarda:2023dqn}, for Drell-Yan production.
While the nature of beyond the Standard Model (BSM) searches is to set limits, the development of next-to-leading logarithmic (NLL) calculations also for BSM searches~\cite{Bozzi:2006fw,Bozzi:2007qr,Bozzi:2007tea}, have been crucial in order to reduce the error from missing higher orders and make full use of the power of the LHC. 

For the hypothetical Drell-Yan like production of scalar leptons, sleptons, in supersymmetric models, the state-of-the-art prediction today is with approximate next-to-next-to-leading order (aNNLO) fixed order contributions, plus next-to-next-to-leading logarithmic resummed results (NNLL) 
\cite{Bozzi:2006fw,Bozzi:2007qr,Bozzi:2007tea,Fiaschi:2018xdm,Fiaschi:2019zgh},
which can be found implemented in the \textsc{Resummino} code \cite{Fiaschi:2023tkq}.\footnote{We will return to a more detailed discussion of what is included here in Section~\ref{sec:existing_calculations}.} A calculation to N$^3$LL has also been made using Soft Collinear Effective Theory (SCET)~\cite{Broggio:2011bd}.

Historically, the main focus in threshold resummation has been on \emph{leading power} (LP) effects. For Drell-Yan like processes at a partonic center-of-mass energy $\sqrt{\hat{s}}$, denoting the invariant mass of the color-singlet part of the final state by $Q^2$, the perturbative expansion of the partonic differential cross-section with resummed contributions can be written schematically in the threshold expansion  as 

\begin{align}
    \frac{d\hat\sigma}{dQ^2} = \sigma_0\sum_{n=0}^\infty \left(\frac{\alpha_s}{\pi}\right)^n&\left[c_n^{\left(\delta\right)}\delta\left(1-z\right)\vphantom{\sum_{m=0}^{2n-1}}\right.\notag\\
    &\left.+\sum_{m=0}^{2n-1}\left[c_{nm}^{\left(-1\right)}\left(\frac{\log^m{\left(1-z\right)}}{1-z}\right)_++c_{nm}^{\left(0\right)}\log^m{\left(1-z\right)}+\order{1-z}\right]\right],
    \label{eq:threshold_expansion}
\end{align}
with $z\equiv Q^2/\hat{s}$ the threshold variable. Here the terms in the inner sum are arranged by powers of $\left(1-z\right)$, which goes to zero in the threshold limit; the $c_{nm}^{\left(-1\right)}$ terms are the most studied LP logarithms, the $c_{nm}^{\left(0\right)}$ terms are \emph{next-to-leading power} (NLP) logarithms, \emph{etc.} The leading logarithmic (LL) terms in the expansion are then the $m=2n-1$ terms, the next-to-leading logarithmic the $m=2n-2$ terms, \emph{etc.}\footnote{As an explicit example, an $\order{\alpha_s}$ calculation will contain leading-power terms proportional to $\alpha_s\left(\frac{\log{\left(1-z\right)}}{1-z}\right)_+$ (leading logarithm) and $\alpha_s\left(\frac{1}{1-z}\right)_+$ (next-to-leading logarithm), and next-to-leading power terms proportional to $\alpha_s\log{\left(1-z\right)}$ (leading logarithm).} The resummation of LL next-to-leading power effects, \emph{i.e.}\ those terms containing $c_{n\left(2n-1\right)}^{\left(0\right)}$, has recently been performed for the SM Drell-Yan process in Refs.~\cite{Bahjat-Abbas:2019fqa,Beneke:2018gvs} for the $q\overline{q}$ channel, and in Ref.~\cite{vanBeekveld:2021mxn} for the $qg$ (and $\overline{q}g$) channel.

Threshold resummation provides a systematic way to control the logarithms that become large when the partonic subprocess approaches threshold, $z\rightarrow 1$, by reorganizing the perturbative series for the partonic cross-section in this limit. Resummation improves the robustness of truncated fixed-order perturbative  predictions, for instance through reduced scale dependence and reduced impact of threshold-enhanced higher-order terms. However, it should not be interpreted as a statement about improving convergence of the perturbative expansion in general. The full perturbative expansion is generally expected to be asymptotic (coefficients exhibiting factorial growth), and \emph{resummation} here is not meant in the mathematical sense of a summation method (e.g.\ Borel summation) that, when applicable, assigns a value to a divergent asymptotic series via a convergent transform/integral, without making the original power series converge. So threshold resummation is an all-order control of a specific subset of terms (the threshold-enhanced ones), not an all-order calculation of the whole cross section. To quantify the perturbative stabilization we present the results through additive \emph{matching}: the fixed-order prediction is combined with the resummed contribution in such a way that no terms are double counted, and the matched result reproduces the fixed-order cross section through the chosen order, while incorporating the resummation-implied logarithmic towers beyond that order. This procedure ensures that resummation modifies the prediction only by higher-order threshold-enhanced contributions, and that away from the threshold-dominated region the matched result reduces to the fixed-order one (up to higher-order effects).

At hadron colliders the partonic threshold limit is not imposed directly, since the observable cross section is obtained after convolution with parton distribution functions. The numerical importance of the near-threshold region is therefore governed by how strongly the convolution weights configurations with $z\simeq 1$. This weighting is controlled by the hadronic variable $\tau=Q^2/s=z \hat s /s= zx_1x_2$, where
$s$ is the hadronic collider energy squared, and $x_i$ are the momentum fractions of each of the colliding partons. For heavy new particles at fixed collider energy, $\tau$ is not very small and the production process probes comparatively large momentum fractions of the partons. Since parton distribution functions are rapidly decreasing as $x\rightarrow1$, contributions far from threshold where $z$ is small, corresponding to larger $x_1 x_2=\tau/z$, are strongly suppressed, and the convolution becomes more concentrated near $z\simeq 1$. For much lighter BSM particles, $\tau\ll 1$, the typical momentum fractions are smaller, the parton distribution functions are less steep, and a broader range of $z < 1$ contribute, reducing the numerical impact of threshold resummation and the matched resummed prediction approaches the fixed-order result, as expected. This convolution-driven enhancement (often referred to as ``dynamical threshold enhancement'') has been quantified for Drell--Yan production in \cite{Becher:2007ty}, to which we refer for a more detailed discussion.

In the present study we will explore the effects of including resummed contributions beyond leading power in BSM cross sections. The different power of $(1-z)$ is expected to affect in particular a more intermediate $\tau$ and particle mass range relative to the hadronic centre-of-mass energy, and we expect such power-suppressed logarithms to play a non-trivial role in the perturbative stability of matched predictions. In particular, we will assess how their inclusion affects the residual scale dependence as a function of the particle mass, and compare this to the improvements obtained by increasing the logarithmic accuracy at leading power alone.

We will review the general methodology in a pedagogical manner with the aim of future extensions, but focus on slepton pair production as a concise BSM example, which has immediate applications to current LHC searches \cite{ATLAS:2019lff,CMS:2020bfa}, and to determining the physics reach of potential future colliders such as the FCC-hh, where the uncertainty from missing higher orders will be more significant at masses currently just outside the reach of the LHC.

We will present complete analytic NLP results and show the numerical effects of including these contributions, matched to fixed-order results, quantifying the remaining cross section uncertainty from higher-order contributions. In doing so we compare to existing results and discuss the impact of the NLP contributions.

The structure of this paper is as follows: in Section~\ref{sec:SUSY_model} we define the supersymmetric model used as our example, while in Section~\ref{sec:NLO} we briefly revise the fixed order calculation for slepton pair production up to next-to-leading order, to clarify our notation and conventions.
Following this, in Section~\ref{sec:resummation}, we review threshold resummation at leading and next-to-leading power, before giving our results for the specific case of slepton pair production in Section~\ref{sec:slepton_resummation}.
We go on to discuss the numerical impact of our results in Section~\ref{sec:results}, comparing to existing results in the literature, before we conclude in Section~\ref{sec:conclusion}. In the appendices we collect the full expressions for the analytical results used in Appendix~\ref{sec:expressions}, and give a, hopefully, pedagogical introduction to the convergence properties of the Mellin transform and its inverse in Appendix~\ref{sec:Mellin transform}.

\section{Supersymmetric model}
\label{sec:SUSY_model}
We use the production of scalar partners to the Standard Model leptons and neutrinos, the sleptons and sneutrinos, as our BSM example. To simplify language we will often refer to both types of particles as sleptons. We perform our calculations within the Minimal Supersymmetric Standard Model (MSSM)~\cite{Nilles:1983ge,Haber:1984rc}. 

The scalar leptons come in three generations, with two gauge eigenstates per generation for the charged sleptons, referred to as the left and right states. In the most general parameterisation of the MSSM, with arbitrary flavour mixing in the soft mass terms, all sleptons with the same electric charge may mix. Here, to simplify notation, we will assume no mixing between generations, and for the first two generations, the selectrons and the smuons, the left and right states will be unmixed, while we will allow mixing for the third generation staus due to the potential significant contribution to off-diagonal terms in the stau mass matrix from the tau Yukawa coupling. Identical assumptions are made for the mass eigenstates of the scalar quarks, squarks.

We write the mixing matrix for the mass eigenstates of the sfermions $\tilde{f}^k_i$, $i=1,2$, in terms of the gauge eigenstates $\tilde{f}^k_{L/R}$, where $k$ indicates the fermion generation,  as
\begin{equation}
    \begin{pmatrix}
        \tilde{f}^k_1\\\tilde{f}^k_2
    \end{pmatrix} = \mathbf{f}^k
    \begin{pmatrix}
        \tilde{f}^k_L\\\tilde{f}^k_R
    \end{pmatrix}\text{.}
    \label{eq:slepton_mixing_matrix}
\end{equation}
Here $f=\ell,q,l,\nu,u,d$, where we will use $\ell$ and $q$ to represent a generic slepton or squark, while $l,\nu,u,d$ indicate a specific type, and $\boldsymbol{\nu}^k_{AB}=\delta_{A1}\delta_{B1}$ as the MSSM contains only left-handed sneutrinos. We also assume that all contributions to the sfermion mass matrices are real so that the mixing matrices are real (and orthogonal) as well.

The numerical results presented below only depend on these assumptions to a very minor degree. Results for full flavor mixing and complex phases in the Lagrangian parameters can be relatively easily derived by generalising the results for the third generation sleptons, substituting the appropriate mixing matrices.

\section{Slepton pair production at NLO}
\label{sec:NLO}
In this section we will briefly outline the computation of the fixed order cross-section for inclusive slepton pair production in proton--proton collisions up to next-to-leading order (NLO), in order to establish and explain some choices of convention. These results have been derived previously in various references, e.g.\ see \cite{Altarelli:1979ub,Dawson:1983fw,delAguila:1990yw,Baer:1993ew,Baer:1997nh,Beenakker:1999xh,Bozzi:2004qq,Bozzi:2007qr}. For detailed expressions for the results we refer to Appendix \ref{sec:expressions}.

\subsection{Leading order}
Proton--proton cross-sections can generically be written as a convolution of the partonic cross-section and parton distribution functions $f_i$ for the proton, with a sum over allowed initial-state parton configurations. Writing the cross-section differential in invariant mass squared of the produced particles, $Q^2$, we have
\begin{align}
    \frac{d\sigma}{dQ^2} 
    &= \sigma_B \sum_{ij}\int_\tau^1\frac{dx_1}{x_1}\int_{\tau/x_1}^1\frac{dx_2}{x_2}f_i\left(x_1,\mu_F\right)f_j\left(x_2,\mu_F\right)w_{ij}\left(z,Q^2,\mu_R,\mu_F\right)\text{,}
    \label{eq:cs_PDFintegral}
\end{align}
where $\mu_R$ and $\mu_F$ are the renormalization and factorization scales, respectively, and $\tau=Q^2/s$ with $\sqrt{s}$ the proton center-of-mass energy. The partonic cross-section has been normalized by the factor $\sigma_B/\left(x_1x_2\right)$, where $\sigma_B$ is the Born cross-section, so that the remaining \emph{partonic coefficient function} $w_{ij}$ only depends on $x_1$ and $x_2$ through the threshold variable $z=\frac{\tau}{x_1x_2}$.

\begin{figure}[ht]
    \centering
    \includegraphics{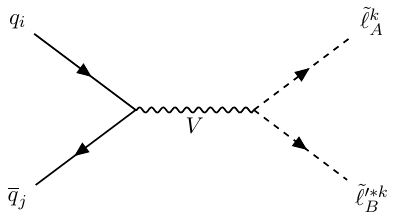}
    \caption{Partonic tree-level diagram for the production of the slepton pair $\tilde{\ell}^k_{A}\tilde{\ell}'^{*k}_{B}$; $V$ can be a photon, $Z$ or $W^\pm$ depending on the final state.}
    \label{fig:treelevel}
\end{figure}

For the production of a pair of sleptons $\tilde{\ell}^k_{A}\tilde{\ell}^{\prime k*}_{B}$, as shown in Figure~\ref{fig:treelevel}, the Born cross section  goes as $\order{\alpha^2}$ and is given by
\begin{equation}
\sigma_B = \frac{\pi\alpha^2\beta^3}{3N_CsQ^2},
\end{equation}
where 
\begin{equation}
\beta^2=1-2\frac{m_{\tilde{\ell}^k_A}^2+m_{\tilde{\ell}'^k_B}^2}{Q^2}+\frac{\left(m_{\tilde{\ell}^k_A}^2-m_{\tilde{\ell}'^k_B}^2\right)^2}{Q^4}.
\end{equation}
In the case of equal final-state slepton masses, $\beta$ is the velocity of the sleptons in their common center-of-mass frame. The tree-level partonic coefficient function is given by
\begin{equation}
    w_{q_i\overline{q}_j}^{\text{LO}}\left(z,Q^2,\mu_R,\mu_F\right) = F^{\tilde{\ell}^k_A\tilde{\ell}'^k_B}_{q_iq_j}\left(Q^2\right)\delta\left(1-z\right)\text{,}
    \label{eq:partonic_coefficient_function_LO}
\end{equation}
where $F^{\tilde{\ell}^k_A\tilde{\ell}'^k_B}_{q_iq_j}\left(Q^2\right)$ is an ``effective coupling'' factor collecting the couplings of the initial-state quarks and final-state sleptons to the intermediate-state boson; its full expression is given in Appendix \ref{sec:expressions}.

\subsection{Next-to-leading order}
\begin{figure}[ht]
    \centering
    \includegraphics{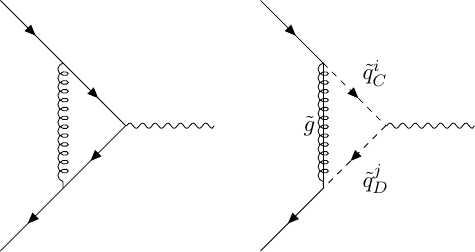}
    \caption{$\order{\alpha_s}$ loop corrections to the quark--vector-boson vertex. Left: Virtual gluon exchange. Right: Gluino--squark loop contribution. $C$ and $D$ denote the squark mass eigenstates.}
    \label{fig:virtual}
\end{figure}

At the next-to-leading order the cross-section gets dominant corrections of $\order{\alpha_s}$. These include virtual corrections to the quark--vector-boson vertex, as shown in Figure~\ref{fig:virtual}, both from gluon exchange and through a squark--gluino loop, which contribute through interference with the tree-level diagram. In the limit of massless quarks the former can be evaluated directly relatively easily, while we express the latter in terms of Passarino-Veltman~\cite{Passarino:1978jh} coefficients. The squark-gluino contribution also gives a correction to the effective coupling, since the vector boson--quark vertex is replaced by a vector boson--squark vertex. The full results are found in Appendix \ref{sec:expressions} in terms of the partonic coefficient function $w_{q_i \bar q_j}^\text{NLO}$.

\begin{figure}[ht]
    \centering
    \begin{subfigure}[c]{0.48\textwidth}
        \includegraphics{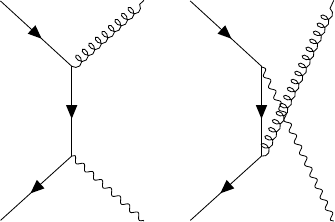}
        \caption{$q_i\overline{q}_j$ initial state.}
        \label{fig:gluon_emission}
    \end{subfigure}
    \begin{subfigure}[c]{0.48\textwidth}
        \includegraphics{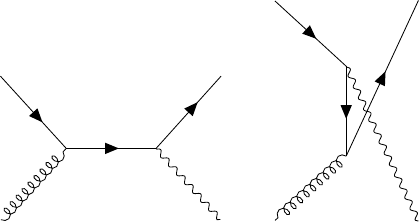}
        \caption{$q_ig$ initial state.}
        \label{fig:quark_emission}
    \end{subfigure}
    \caption{Real-emission diagrams that contribute to $\order{\alpha_s}$, for the different initial states that contribute to this order. The diagram for the $\overline{q}_ig$ initial state is obtained by simply reversing the fermion arrow in the $q_ig$ diagram.}
    \label{fig:real}
\end{figure}

We also have contributions from real gluon emission off the incoming quarks, shown in Figure~\ref{fig:gluon_emission}, giving another $\order{\alpha_s}$ contribution to the $q_i\overline{q}_j$ channel; and real quark emission diagrams shown in Figure~\ref{fig:quark_emission}, which constitute the lowest-order contribution to the $q_ig$ and $\overline{q}_ig$ channels. Again, we refer to Appendix \ref{sec:expressions} for expressions for these contributions in terms of $w_{q_i \bar q_j}^\text{NLO}$, $w_{gq_i }^\text{NLO}$ and $w_{g\bar q_i }^\text{NLO}$.

\subsection{Renormalization}
The diagrams listed above give rise to various divergences; the way these are handled depends on their origin. 

\begin{figure}[ht]
    \centering
    \includegraphics{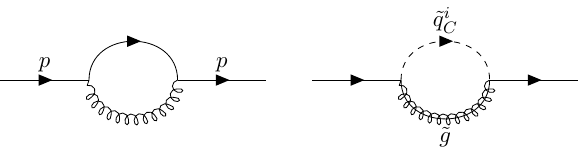}
    \caption{$\order{\alpha_s}$ contributions to the quark (of flavor $i$) self-energy. As in Figure~\ref{fig:virtual} there is an additional sum over the squark mass eigenstate $C$.}
    \label{fig:counterterms}
\end{figure}

The virtual diagrams of Figure~\ref{fig:virtual} contain ultraviolet divergences, which are removed through on-shell renormalization of the external quark field(s), {\it e.g.}\ see Ref.~\cite{Altarelli:1979ub}. In practice this amounts to adding the counterterm\footnote{The quarks are here taken to be massless.}
\begin{equation}
    \delta_2=\left.-\frac{1}{2}\frac{d}{d\slash{p}}\left(\Sigma^i_2\left(\slash{p}\right)+\Sigma^j_2\left(\slash{p}\right)\right)\right\rvert_{\slash{p}=0}\text{,}
\end{equation}
with the $\order{\alpha_s}$ contributions to the quark self-energy $i\Sigma_2\left(\slash{p}\right)$ shown in Figure~\ref{fig:counterterms}. Note that we get contributions from each of the two (potentially distinct) initial-state quark flavors $i$ and $j$. Similarly to the virtual corrections, these contribute through interference with the tree-level diagram. Again, the Standard Model corrections can be expressed in closed form---as it is proportional to the scaleless integral $\int\frac{d^dk}{k^4}$, its only effect is to replace a UV divergence by an IR one. The squark--gluino loop result also has a finite component, which is given in terms of Passarino--Veltman coefficients and also results in a changed effective coupling. The results can be found in Appendix \ref{sec:expressions}, in terms of $w^{\text{NLO}}_{q_i\overline{q}_j}$.

Both the real and virtual diagrams contain infrared divergences, most of which are canceled when the two are added. There still remains collinear divergences, however, in both the $q\overline{q}$ and $qg$ channels these are absorbed by the PDFs through $\MSbar$ renormalization.

To summarize our choices of renormalization schemes, we use on-shell renormalization of all masses, decay widths and external fields, while the strong coupling $\alpha_s$ and the PDFs are renormalized in the $\MSbar$ scheme.

The electroweak quantities $\alpha$ and $\theta_W$ are set by the $G_\mu$ scheme~\cite{Denner:1991kt,Dittmaier:2001ay}, in which the pole masses $m_W$ and $m_Z$ (plus decay rates), and the Fermi constant $G_F$, are taken as input parameters (and defined in the on-shell scheme). 
The widths of the $W$ and $Z$ bosons are included using the \emph{complex mass scheme} to preserve gauge invariance~\cite{Denner:1999gp,Denner:2005fg,Denner:2006ic}, in which the vector boson masses are replaced by the (complex) location of the pole in the Breit-Wigner propagator: 
\begin{equation}
    m_V^2 \rightarrow \mu_V^2 \equiv m_V^2-im_V\Gamma_V\text{,}
    \label{eq:cms}
\end{equation}
where $\Gamma_V$ is the decay width of the boson in question, fixed by measurements. This replacement is also made in defining the Weinberg angle as
\begin{equation}
    \sin\theta_W = 1-\frac{\mu_W^2}{\mu_Z^2}\text{,}
    \label{eq:sintheta_cms}
\end{equation}
while $\alpha$ is given by
\begin{equation}
    \alpha = \sqrt{2}\frac{G_Fm_W^2}{\pi}\left(1-\frac{m_W^2}{m_Z^2}\right)\text{.}
\end{equation}
Note that the latter uses the real $W$ and $Z$ masses to avoid spurious terms of $\mathcal O (\alpha)$~\cite{Denner:2019vbn}.

\subsection{Threshold limit}
\label{sec:fixed_threshold}
While the full expressions for the NLO cross-section are not too instructive on their own, their behavior in the threshold limit of $z\rightarrow 1$ help illustrate the need for resummation. Taking this limit in the next-to-leading order partonic coefficient functions $w_{ij}^{\text{NLO}}$, \eqref{eq:qq_NLO} and \eqref{eq:qg_NLO}, retaining only the dominant terms, and leaving out the terms proportional to $\delta\left(1-z\right)$, we find the threshold behavior,\footnote{Here, and in the following, $\alpha_s\equiv\alpha_s\left(\mu_R\right)$ unless otherwise stated.}
\begin{align}
    w_{q_i\overline{q}_j}^{\text{NLO}}\left(z\right) &\simeq F^{\tilde{\ell}^k_A\tilde{\ell}'^k_B}_{q_iq_j}\left(Q^2\right)\frac{C_F\alpha_s}{\pi}\left\{4\left(\frac{\log{\left(1-z\right)}}{1-z}\right)_+-2\log{\frac{\mu_F^2}{Q^2}}\frac{1}{\left(1-z\right)_+}-4\log{\left(1-z\right)}\right\}\text{,}
    \label{eq:qq_threshold}\\
    w_{q_ig}^{\text{NLO}}\left(z\right) &\simeq \sum_jF^{\tilde{\ell}^k_A\tilde{\ell}'^k_B}_{q_iq_j}\left(Q^2\right)\frac{T_F\alpha_s}{\pi}\log{\left(1-z\right)}\text{,}
    \label{eq:qg_threshold}
\end{align}
where we have suppressed all arguments of $w_{ij}$ except for $z$, and we will continue to do so below. 

Alternatively, we can take the Mellin transform of the partonic coefficient function. Denoting the Mellin transform of a function $f(z)$ by the same symbol capitalized $F(N)$, the transform and its inverse is defined as (see Appendix~\ref{eq:analytical_properties_Mellin} for more detail)
\begin{align}
    F\left(N\right) &= \int_0^1dz\,z^{N-1}f\left(z\right)\text{,}\label{eq:mellin} \\
    f\left(z\right) &= \frac{1}{2\pi i}\int_{c-i\infty}^{c+i\infty}dN\,z^{-N}F\left(N\right)\text{.}\label{eq:invmellin}
\end{align}
Using this definition,
the partonic coefficient functions can be written in the large-$N$ limit as
\begin{align}
    W_{q_i\overline{q}_j}^{\text{NLO}}\left(N\right) &\simeq F^{\tilde{\ell}^k_A\tilde{\ell}'^k_B}_{q_iq_j}\left(Q^2\right)\frac{C_F\alpha_s}{\pi}\left\{2\log^2{\overline{N}}+2\log{\frac{\mu_F^2}{Q^2}}\log{\overline{N}}+2\frac{\log{\overline{N}}}{N}\right\}\text{,}
    \label{eq:qq_threshold_N}\\
    W_{q_ig}^{\text{NLO}}\left(N\right)&\simeq -\sum_jF^{\tilde{\ell}^k_A\tilde{\ell}'^k_B}_{q_iq_j}\left(Q^2\right)\frac{T_F\alpha_s}{\pi}\frac{\log{\overline{N}}}{N}
    \label{eq:qg_threshold_N}\text{,}
\end{align}
where $\overline{N}\equiv Ne^{\gamma_E}$, with $\gamma_E$ being the Euler-Mascheroni constant. 
The first two terms in \eqref{eq:qq_threshold}, or equivalently in \eqref{eq:qq_threshold_N}, are the leading and next-to-leading logarithmic terms from Eq.~(\ref{eq:threshold_expansion}), respectively, at leading power in $\left(1-z\right)$ (or equivalently in $N$). The remaining terms are the leading logarithmic terms at the next-to-leading power.

Though these terms are all integrable in a distributional sense, they may still dominate the resulting cross-section, in particular in the limit of high slepton masses where smaller-$z$ regions are suppressed by the smallness of the PDFs for large momentum fractions. Furthermore, similar large logarithms appear at all orders in perturbation theory, indicating that a pure fixed-order calculation will have limited predictive power as potentially large higher-order corrections are missed. Here, systematic resummation of these terms can help.

\section{Threshold resummation at leading and next-to-leading power}
\label{sec:resummation}
Threshold resummation relies on the factorization and exponentiation properties of QCD in Mellin space. Firstly, amplitudes and cross-sections for a variety of different processes are known to factorize to products of hard and soft effects.\footnote{See, e.g., Ref.~\cite{Collins:1989gx} for a comprehensive review.} Secondly, in the threshold limit, the soft factor exponentiates in terms of Feynman diagrams called \emph{webs}~\cite{Gatheral:1983cz,Frenkel:1984pz}, allowing for these effects to be resummed to all orders in the strong coupling.

Here, we briefly review and summarize the methods for performing this resummation for Drell-Yan like processes in the leading-power (LP) approximation, before outlining how the procedure is generalized to the next-to-leading power (NLP) case. Since we are using resummation mainly as a means to stabilize our fixed-order calculations, we restrict ourselves to resumming leading-power terms up to NLL, and next-to-leading power terms up to LL, as these contain terms that appear at NLO. Higher logarithmic terms would first appear in the NNLO fixed-order calculation, which currently has not been performed for slepton pair production.

\subsection{Leading-power resummation}
\label{sec:LPres}
The $q\overline{q}$ partonic coefficient function for the Drell-Yan process near threshold, can be written in Mellin space to leading power in the factorized form~\cite{Sterman:1986aj,Kidonakis:1997gm} 
\begin{equation}
    W_{q\overline{q}}^{\text{LP}}\left(N\right) \sim \left\lvert H_{q\overline{q}}\left(Q^2\right)\right\rvert^2\frac{\prod_{i=q,\overline{q}}\psi_i\left(N,Q^2\right)}{\prod_{i=q,\overline{q}}\phi_{i}\left(N,Q^2\right)}S_{q\overline{q}}\left(N,Q^2\right)\text{,}
    \label{eq:factorization}
\end{equation}
where we have suppressed some scale arguments that are immaterial to the following discussion.

The $N$-independent\footnote{Contributions proportional to $\delta(1-z)$ in $x$-space.} hard function $H_{q\overline{q}}$ in (\ref{eq:factorization}) contains the high-energy, virtual corrections; the soft function $S_{q\overline{q}}$ collects low-energy effects; and the quark distribution functions $\psi_i$ contain collinear effects. To avoid double-counting of effects that are both soft and collinear --- which are included both in $\psi_i$ and $S_{q\overline{q}}$ --- the quark distributions are divided by the light-cone distributions $\phi_i$. These are the eikonal approximation to the quark distribution functions $\psi_i$, and thus effectively serve to remove this overlap. 
The ``$\sim$'' notation reflects the fact that this factorization formula only holds for the \emph{bare} cross-section, i.e.\ before renormalization; thus there will generally be some regularization dependence in \eqref{eq:factorization} as well.

The resummation of the LP threshold effects to NLL order was shown in Refs.~\cite{Sterman:1986aj,Catani:1989ne}, and has since been verified using a multitude of methods, including using Wilson loops and their renormalization properties~\cite{Korchemsky:1992xv,Korchemsky:1993uz}, and Soft Collinear Effective Theory (SCET)~\cite{Becher:2006nr,Schwartz:2007ib,Bauer:2008dt,Chiu:2009mg}. The exponentiation also relies on the fact that in the soft limit, $n$-gluon phase space integrals factorize into $n$ 1-gluon phase space integrals.

In the end, the partonic coefficient function is given by~\cite{Sterman:1986aj,Catani:1989ne}
\begin{align}
    W^{\text{res,LP}}_{q\overline{q}}\left(N\right) = \left\lvert H_{q\overline{q}}\left(Q^2\right)\right\rvert^2&\exp{\left\{\int_0^1dzz^{N-1}\left[\vphantom{\int_{\mu_F^2}^{\left(1-z\right)^2Q^2}}\frac{1}{1-z}D\left(\alpha_s\left(\left(1-z\right)^2Q^2\right)\right)\right.\right.}\notag\\
    &\quad\qquad\left.\left.+2\int_{\mu_F^2}^{\left(1-z\right)^2Q^2}\frac{dq^2}{q^2}P^{\text{LP}}_{qq}\left(z,\alpha_s\left(q^2\right)\right)\right]_+\right\}\text{,}
    \label{eq:LPres}
\end{align}
where the Mellin transform in the exponential has been left unevaluated.
Here $D\left(\alpha_s(q^2)\right)$ collects soft wide-angle contributions; written as a perturbation series, leaving out the scale dependence of $\alpha_s$ it is given by
\begin{equation}
    D\left(\alpha_s\right) = D_1\frac{\alpha_s}{\pi}+D_2\left(\frac{\alpha_s}{\pi}\right)^2+\dots\text{.}
    \label{eq:resum_D}
\end{equation}
The only coefficient needed for NLL resummation is $D_1$, which is zero. 
Here, $P_{qq}^{\text{LP}}$ is the DGLAP splitting function to leading power in the threshold expansion; explicitly~\cite{Vogt:2004mw,Moch:2004pa}, 
\begin{equation}
    P^{\text{LP}}_{qq}\left(z,\alpha_s\left(q^2\right)\right) = \left(\frac{1}{1-z}\right)_+A\left(\alpha_s\left(q^2\right)\right)\text{,}
    \label{eq:LP_splitting}
\end{equation}
with
\begin{equation}
    A(\alpha_s) = A_1\frac{\alpha_s}{\pi}+A_2\left(\frac{\alpha_s}{\pi}\right)^2+\dots\text{,}
    \label{eq:splitting_A}
\end{equation}
where higher-order terms beyond those shown first contribute at NNLL.

\subsection{Next-to-leading power resummation}
\label{sec:NLPres}
Factorization theorems analogous to \eqref{eq:factorization} have been derived for the Drell-Yan process also at next-to-leading power in the threshold expansion, for both the quark--antiquark~\cite{Beneke:2019oqx} and quark--gluon~\cite{Broggio:2023pbu} initial states, meaning that if resummation of the infrared effects can be achieved, NLP effects can also be included to all orders. As the authors of these articles note, these expressions do not straightforwardly allow for resummation above LL accuracy, but LL resummation can still be achieved.

At next-to-leading power, corrections can in principle arise in two ways; either from NLP effects in the squared amplitude, or from corrections to the phase space. Schematically, we can write these corrections to the cross-section as~\cite{Bahjat-Abbas:2019fqa}
\begin{equation}
    \sigma^{\text{NLP}} \sim \int d\Phi_{\text{LP}}\left\lvert\mathcal{M}\right\rvert_{\text{NLP}}^2+\int d\Phi_{\text{NLP}}\left\lvert\mathcal{M}\right\rvert_{\text{LP}}^2\text{,}
\end{equation}
where $\Phi_{\text{LP}}$ is the leading-power phase space which is known to factorize (see above), while $\Phi_{\text{NLP}}$ contains non-factorizing corrections. These corrections have been shown~\cite{Bahjat-Abbas:2019fqa} to contribute only at subleading logarithmic order; thus, if we restrict ourselves to only a LL resummation of NLP logarithms, this contribution can be neglected. 

As for the NLP corrections to the squared amplitude, they can arise in two different ways: Either as ``next-to-soft'' gluon radiation from the incoming hard partons, but separate from the main hard interaction --- these will result in generalizations of the soft functions discussed above, which similarly exponentiate --- or soft radiation originating from inside the hard interaction itself.  When the external particles are massless, as they are in this case, the latter effects contain collinear singularities, and will not necessarily factorize and exponentiate as cleanly as the former.
These collinear singularities can be controlled by introducing \emph{radiative jet functions,} which are essentially generalizations to the quark distribution functions of \eqref{eq:factorization}. 
These were originally introduced for QED~\cite{DelDuca:1990gz}, and have more recently been extended to QCD as well and computed to one-loop order~\cite{Bonocore:2015esa,Bonocore:2016awd}. 

For the present discussion, radiative jet functions do not have to be taken into account, however; as in \eqref{eq:factorization} the overlap between the radiative jet functions and the next-to-soft contributions must be removed, and to leading order in NLP logarithms this removes any effect of these functions~\cite{Bahjat-Abbas:2019fqa}.

The next-to-soft contributions can be included by considering so-called \emph{generalized Wilson lines;} these were constructed through a path integral approach in Ref.~\cite{Laenen:2008gt}, by allowing for slight corrections to the straight-line trajectories of the incoming quarks. When using the leading-power phase space as discussed above, these contributions exponentiate, which then leads to the combined LP and NLP resummed expression~\cite{Bahjat-Abbas:2019fqa}
\begin{align}
    W^{\text{res,NLP}}_{q\overline{q}}\left(N\right) = \left\lvert H_{q\overline{q}}\left(Q^2\right)\right\rvert^2&\exp{\left\{\int_0^1dzz^{N-1}\left[\frac{1}{1-z}D\left(\alpha_s\left(\frac{\left(1-z\right)^2Q^2}{z}\right)\right)\right.\right.}\notag\\
    &\quad\qquad\left.\left.+2\int_{\mu_F^2}^{\left(1-z\right)^2Q^2/z}\frac{dq^2}{q^2}P^{\text{LP}+\text{NLP}}_{qq}\left(z,\alpha_s\left(q^2\right)\right)\right]_+\right\}\text{.}
    \label{eq:NLPres}
\end{align}

This differs from \eqref{eq:LPres} in two ways; firstly by retaining the full $z$ dependence of the emitted gluon energy in the argument of $D\left(\alpha_s\right)$ and in the upper limit of the $q^2$ integral; secondly, the splitting function is now approximated to next-to-leading power:
\begin{equation}
    P^{\text{LP}+\text{NLP}}_{qq}\left(z,\alpha_s\left(q^2\right)\right) = \left[\left(\frac{1}{1-z}\right)_+-1\right]A\left(\alpha_s\left(q^2\right)\right)\text{.}
    \label{eq:NP_splitting}
\end{equation}
An equivalent result was also found using SCET in Ref.~\cite{Beneke:2018gvs}.

\subsubsection{Resummation of the quark--gluon partonic cross-section}
\label{sec:qgres}
The discussion so far has exclusively been around the color-singlet $q\overline{q}$ initial-state contribution. This type of initial state is historically the most closely studied in resummation literature, as it is the only one to result in LP logarithms, as demonstrated to NLO in \eqref{eq:qq_threshold}--\eqref{eq:qg_threshold}. At NLP there are also logarithmic contributions from the $qg$ ($\overline{q}g$) initial state; to consistently treat all NLP effects, these must also be resummed to all orders. 

The NLP resummation of the $qg$ partonic cross-section is in a sense simpler than that of the $q\overline{q}$ one, due to the fact that baryon number conservation necessitates at least one quark emission. A quark (or any other fermion) in the final state, with the soft momentum $k$, contributes to the squared amplitude a factor $\sum_s u^s\left(k\right)\overline{u}^s\left(k\right) = \slash{k}$, which is a small factor if the emitted quark is soft. The corresponding spin sum for a final-state gluon contains only the spacetime metric $\eta^{\mu\nu}$; thus soft-quark emission diagrams are automatically suppressed by one factor of soft momenta relative to soft-gluon emission. In other words, just a single final-state fermion leaves the cross-section at NLP in soft radiation momentum. Any additional radiation can then be calculated using the standard eikonal approximation, and the factorizable LP phase space.

An all-order diagrammatic calculation using these constraints leads to the following result which is explicitly in Mellin space~\cite{vanBeekveld:2021mxn} (this was conjectured earlier in Ref.~\cite{LoPresti:2014ihe}):
\begin{align}
    &W^{\text{res,NLP}}_{qg}\left(N\right) = \left\lvert H_{qg}\left(Q^2\right)\right\rvert^2\frac{T_F}{C_A-C_F}\frac{1}{2N\log{N}}&\notag\\
    &\quad\times\left[\exp{\left\{\frac{2\alpha_s}{\pi}C_F\log^2{N}\right\}}\mathcal{B}_0\left(\frac{\alpha_s}{\pi}\left(C_A-C_F\right)\log^2{N}\right)-\exp{\left\{\frac{\alpha_s}{2\pi}\left(C_F+3C_A\right)\log^2{N}\right\}}\right]\text{.}
    \label{eq:qgres}
\end{align}
The function $\mathcal{B}_0\left(\frac{\alpha_s}{\pi}\left(C_A-C_F\right)\log^2{N}\right)$ originates from the LL approximation to the splitting function $P_{qg}\left(N\right)$~\cite{Vogt:2010cv}, and is given by
\begin{equation}
    \mathcal{B}_0\left(x\right) \equiv 1-\frac{x}{2}-\sum_{n=1}^\infty\frac{\left(-1\right)^n}{\left(2n!\right)^2}\left\lvert B_{2n}\right\rvert x^{2n}\text{,}
\end{equation}
where $B_n$ are the Bernoulli numbers.

\section{Slepton pair production cross-section at next-to-leading power}
\label{sec:slepton_resummation}
Having outlined the way NLP threshold logarithms can be resummed, we now turn to the specific process of slepton pair production. We begin by finding the expressions for the partonic coefficient functions  directly from the results in the previous section and knowledge of the fixed-order process. We will then discuss matching of these contributions to the fixed-order result, and some technicalities in the numerical evaluation of the resummed result when it is added to the fixed-order results of Section~\ref{sec:NLO}.

\subsection{Resummed result with next-to-leading power logarithms}
\label{sec:slepton_nlpresult}
For the quark--antiquark initial state, with the hard function being the effective coupling from (\ref{eq:partonic_coefficient_function_LO}), we have from \eqref{eq:NLPres} that the resummed contribution to the cross-section to next-to-leading power can be written implicitly in Mellin space as the exponentiated integral
\begin{align}
    W_{q_i\overline{q}_j}^\text{res}\left(N\right) = F_{q_iq_j}^{\tilde{\ell}^k_A\tilde{\ell}'^k_B}\left(Q^2\right)&\exp{\left\{\int_0^1dzz^{N-1}\left[\frac{1}{1-z}D\left(\alpha_s\left(\frac{\left(1-z\right)^2Q^2}{z}\right)\right)\right.\right.}\notag\\
    &\quad\qquad\left.\left.+2\int_{\mu_F^2}^{\left(1-z\right)^2Q^2/z}\frac{dq^2}{q^2}P^{\text{LP}+\text{NLP}}_{qq}\left(z,\alpha_s\left(q^2\right)\right)\right]_+\right\}\text{.}
    \label{eq:qqres}
\end{align}
In addition to the functions discussed in the previous section, we need the $\beta$-function for $\alpha_s$ to evaluate the scale dependence of the different parts of the integrand:
\begin{equation}
    \mu\frac{d\alpha_s}{d\mu} = -2\alpha_s\left(\alpha_sb_0+\alpha_s^2b_1+\dots\right)\text{.}
    \label{eq:alphas_beta}
\end{equation}
The one-~\cite{Gross:1973id,Politzer:1973fx} and two-loop~\cite{Caswell:1974gg,Jones:1974mm,Egorian:1978zx} coefficients are sufficient to find $\alpha_s$ to allow for leading-power resummation up to NLL, and next-to-leading power resummation to LL. In solving this $\beta$-function scale dependence on the reference scale $\mu_R$ appears.

The evaluation of the $q^2$ integral in \eqref{eq:qqres} will also contain some next-to-leading power terms beyond the leading logarithms; we will discard these as they do not constitute the full NLL next-to-leading power result, meaning that this would be an incomplete resummation of these effects. This was discussed in Ref.~\cite{Laenen:2008ux}, where an equation on the form \eqref{eq:qqres} was postulated, and its $\order{\alpha_s^2}$ NLL next-to-leading power logarithms was compared to the fixed order NNLO result, and it was demonstrated that additional NLL next-to-leading power terms exist. As noted in Ref.~\cite{Bahjat-Abbas:2019fqa}, this indicates the need for the radiative jet functions mentioned in Section~\ref{sec:NLPres}, along with corrections to the leading power phase space, as these are both potential next-to-leading power effects that begin to contribute to NLL order.

Note that in \eqref{eq:qqres}, we have not included any terms of $\order{\alpha_s}$ in the hard factor in the cross-section, as these are not needed for NLL resummation. In multiplication with the exponentiated factor they would produce NNLL leading power terms, as well as NLL next-to-leading power terms, which are not resummed to all orders here, nor do they appear in the NLO fixed-order cross-section.

With this, integrating over $q^2$ using the $\beta$-function of $\alpha_s$, and performing the Mellin space transform, the $q\overline{q}$ resummed partonic cross-section can be written in terms of the different logarithms that we do keep on the form~\cite{vanBeekveld:2021hhv}
\begin{align}
    W_{q_i\overline{q}_j}^\text{res}\left(N\right) &= F_{q_iq_j}^{\tilde{\ell}^k_A\tilde{\ell}'^k_B}\left(Q^2\right)\exp{\left\{\left(1+\frac{1}{2}\frac{\partial}{\partial N}\right)\log{\overline{N}}g_1\left(\lambda\right)+g_2\left(\lambda\right)\right\}}\\
    &\equiv F_{q_iq_j}^{\tilde{\ell}^k_A\tilde{\ell}'^k_B}\left(Q^2\right)\exp{\left\{\log{\overline{N}}g_1\left(\lambda\right)+g_2\left(\lambda\right)+h_1\left(\lambda,N\right)\right\}}\text{,}
    \label{eq:g1g2h1}
\end{align}
where we use $\overline{N}\equiv Ne^{\gamma_E}$, with $\gamma_E$ being the Euler-Mascheroni constant, and $\lambda \equiv b_0\alpha_s\log{\overline{N}}$. Here, $\log{\overline{N}}g_1$ and $g_2$ contain the leading and next-to-leading logarithms, respectively, at leading power, while $h_1$ contains leading logarithms at next-to-leading power. The full expressions for the logarithms are listed in Appendix~\ref{sec:expressions} as (\ref{eq:g1})--(\ref{eq:h1}).

As the authors of Ref.~\cite{Bahjat-Abbas:2019fqa} point out, exponentiation of the NLP function $h_1\left(\lambda,N\right)$ technically leads to the inclusion of some higher-power terms as well: The Taylor expansion of the exponential in \eqref{eq:g1g2h1} will contain terms of the form $\log^m{\overline{N}}/N^n$, which is formally N$^n$LP. Since the exponentiation of the NLP soft function on this form has been demonstrated in multiple works~\cite{Laenen:2008gt,Laenen:2010uz,Bahjat-Abbas:2019fqa}, this is more an issue of nomenclature than a significant problem; we still refer to the result as a NLP resummed result, similar to Ref.~\cite{vanBeekveld:2021hhv}, since the exponentiated webs, are at next-to-leading power. This differs from the effects of higher logarithmic order that were left out, which were discussed above.

It is possible to retain only the NLP effects by approximating $\exp{\left(h_1\left(\lambda,N\right)\right)}\simeq 1+h_1\left(\lambda,N\right)$~\cite{Bahjat-Abbas:2019fqa}. We have verified numerically that this only shifts the cross-section by a relative $\order{10^{-3}}$; for this reason, and as discussed above, we opt not to do so (again, similarly to Ref.~\cite{vanBeekveld:2021hhv}).

Turning to the quark--gluon initial state, its resummed contribution to the cross-section is given from \eqref{eq:qgres} by
\begin{align}
    W^{\text{res}}_{q_ig}&\left(N\right) = \sum_jF^{\tilde{\ell}^k_A\tilde{\ell}'^k_B}_{q_iq_j}\left(Q^2\right)\frac{T_F}{C_A-C_F}\frac{1}{2N\log{N}}\notag\\
    &\times\left[\exp{\left\{\frac{2\alpha_s}{\pi}C_F\log^2{N}\right\}}\mathcal{B}_0\left(\frac{\alpha_s}{\pi}\left(C_A-C_F\right)\log^2{N}\right)-\exp{\left\{\frac{\alpha_s}{2\pi}\left(C_F+3C_A\right)\log^2{N}\right\}}\right]\text{.}
    \label{eq:qgres_slepton}
\end{align}
For the hard function the same arguments as above for the quark--antiquark initial state apply. In evaluating the $\mathcal{B}_0$ function numerically, we truncate the sum when the norm of the latest term is less than $10^{-3}$ that of the total function.

\subsection{Matching to fixed-order calculation}
\label{sec:matching}
Care is required in adding these resummed results to the fixed-order results, as there is some overlap. The goal of the matching procedure is to construct a single prediction that \begin{enumerate}
\item is exactly fixed-order accurate up to the order we claim, 
\item is improved near threshold by adding the all-order towers of threshold-enhanced logarithms, and
\item avoids double counting between the fixed-order and resummed contributions. 
\end{enumerate}

As discussed in Section~\ref{sec:fixed_threshold}, the NLO results contain threshold logarithms at $\order{\alpha_s}$, while the resummed results contain these to all orders. To avoid including these terms twice, we subtract from the resummed expressions their Taylor expansion up to $\order{\alpha_s}$. Explicitly, the complete partonic coefficient function is
\begin{equation}
    w_{ij}\left(z\right) = w^{\text{LO}}_{ij}\left(z\right)+w^{\text{NLO}}_{ij}\left(z\right)+\frac{1}{2\pi i}\int_{c-i\infty}^{c+i\infty}dNz^{-N}\left(W^{\text{res}}_{ij}\left(N\right)-W^{\text{exp}}_{ij}\left(N\right)\right)\text{,}
    \label{eq:matched_cs}
\end{equation}
where an inverse Mellin transform is performed on the resummed contributions to move them from Mellin space to the $z$-dependence of the fixed order result. The ``expanded'' partonic functions that are subtracted off are given by
\begin{align}
    W^{\text{exp}}_{q_i\overline{q}_j}\left(N\right) &= F_{q_iq_j}^{\tilde{\ell}^k_A\tilde{\ell}'^k_B}\left(Q^2\right)\left[1+C_F\frac{\alpha_s}{\pi}\left(2\log^2{\overline{N}}+2\log{\frac{\mu_F^2}{Q^2}}\log{\overline{N}}+2\frac{\log{\overline{N}}}{N}\right)\right]\text{,}\\
    W^{\text{exp}}_{q_ig}\left(N\right) &= -\sum_jF^{\tilde{\ell}^k_A\tilde{\ell}'^k_B}_{q_iq_j}\left(Q^2\right)T_F\frac{\alpha_s}{\pi}\frac{\log{N}}{N}\text{.}
    \label{eq:qg_exp}
\end{align}
We recognize these expressions as the large-$N$ limit of the NLO result, which we gave in \eqref{eq:qq_threshold_N}--\eqref{eq:qg_threshold_N}.\footnote{The remaining difference between \eqref{eq:qg_threshold_N} and \eqref{eq:qg_exp} effectively amounts to different approximations in the large-N limit of the Mellin transform $\int_0^1dx\,x^{N-1}\log{\left(1-x\right)}=-\left(\psi\left(N+1\right)+\gamma_E\right)/N$, where the digamma function $\psi\left(1+N\right)\rightarrow \log{N}$ asymptotically for large $N$. The $qg$ resummed result in \eqref{eq:qgres_slepton} follows Ref.~\cite{vanBeekveld:2021mxn} in not including $\gamma_E$ in this approximation, while in \eqref{eq:qg_threshold_N} it was included for symmetry with the $q\overline{q}$ result. In the matching procedure, only the overlap between the resummed and fixed-order results is subtracted, which is exactly \eqref{eq:qg_exp}.}

\subsection{Numerical convergence of the inverse Mellin transform}
\label{sec:inverse_Mellin}
To add the resummed contribution to the cross-section in \eqref{eq:matched_cs} we must perform an inverse Mellin transform numerically. 
Using the definition from \eqref{eq:mellin} on the PDF integral in \eqref{eq:cs_PDFintegral}, we have the very concise result
\begin{equation}
    \frac{d\Sigma}{dQ^2}\left(N\right) = \sigma_B\sum_{ij}F_i\left(N,\mu_F\right)F_j\left(N,\mu_F\right)W_{ij}\left(N,Q^2,\mu_R,\mu_F\right)\text{.}
    \label{eq:cs_PDFintegral_Mellin}
\end{equation}
This means that if the PDFs were available at the same convenience and precision in Mellin space as in physical space --- which they are not, at the time of writing --- the evaluation of the cross-section could be reduced to a single numerical integral, namely an overall inverse Mellin transform, instead of the inverse Mellin transform of the resummed contribution followed by the PDF integrals.

However, the simple factorized form of \eqref{eq:cs_PDFintegral_Mellin} can still be used to stabilize the numerical evaluation. Following Ref.~\cite{Kulesza:2002rh}, since $W_{ij}$ can potentially oscillate significantly for large values of $N$, this can be alleviated by defining the differentiated PDFs
\begin{equation}
    \tilde{f}\left(x,\mu_F\right) \equiv-x\frac{df\left(x,\mu_F\right)}{dx}\text{,}
\end{equation}
which have the property that
\begin{equation}
    \tilde{F}\left(N,\mu_F\right) = -\int_0^1dxx^N\frac{df\left(x,\mu_F\right)}{dx} = NF\left(N,\mu_F\right)\text{,}
\end{equation}
using integration by parts.

This can in turn be used to define the function 
\begin{equation}
    \tilde{W}_{ij}\left(N,Q^2,\mu_R,\mu_F\right) \equiv \frac{W_{ij}\left(N,Q^2,\mu_R,\mu_F\right)}{N^2}\text{,}
\end{equation}
which should be more numerically stable in the large-$N$ limit, writing instead
\begin{equation}
    \frac{d\Sigma}{dQ^2}\left(N\right) = \sigma_B\sum_{ij}\tilde F_i\left(N,\mu_F\right)\tilde F_j\left(N,\mu_F\right)\tilde W_{ij}\left(N,Q^2,\mu_R,\mu_F\right)\text{.}
\end{equation}
This allows us to rewrite \eqref{eq:cs_PDFintegral} as
\begin{equation}
    \frac{d\sigma}{dQ^2} = \sigma_B \sum_{ij}\int_0^1\frac{dx_1}{x_1}\int_0^1\frac{dx_2}{x_2}\tilde{f}_i\left(x_1,\mu_F\right)\tilde{f}_j\left(x_2,\mu_F\right)\tilde{w}_{ij}\left(z=\frac{\tau}{x_1x_2},Q^2,\mu_R,\mu_F\right)\text{,}
    \label{eq:cs_PDFintegral_tilde}
\end{equation}
where the lower limit on the integrals is due to the fact that the inverse Mellin transform of $W_{ij}$ is not necessarily zero for $z>1$, which allows for $x$-values down to zero~\cite{Catani:1996yz}. We discuss this issue, which is related to the convergence of the Mellin transform, in more detail in Appendix~\ref{sec:convergence_properties_Mellin}. Equation (\ref{eq:cs_PDFintegral_tilde}) is then what we use to evaluate the contribution from the resummed terms in (\ref{eq:matched_cs}) to the total cross section.

The differentiated PDFs used in \eqref{eq:cs_PDFintegral_tilde} are based on standard PDFs grid using the \textsc{LHAPDF} package~\cite{Buckley:2014ana}. To speed up the numerical evaluation, the differentiated PDFs are pre-calculated and stored as individual PDF sets, on similar grids as the ordinary \textsc{LHAPDF} grids.

To numerically perform the inverse Mellin transform and find the relevant $\tilde w_{ij}$ contributions in (\ref{eq:cs_PDFintegral_tilde}), we use the \emph{Minimal Prescription} of Ref.~\cite{Catani:1996yz}. For our purposes this amounts to choosing $c=C_{\text{MP}}$ in (\ref{eq:matched_cs}), where $C_{\text{MP}}$ lies on the real axis between the rightmost singularity in the PDFs, and the Landau pole at $\log{\overline{N}}=1/\left(2\alpha_sb_0\right)$.\footnote{The existence of the Landau pole can be seen directly from the $\log{\left(1-2\lambda\right)}$ dependence in the resummed contribution to the cross-section, see \eqref{eq:qqres_app}--\eqref{eq:h1}. We must take into account the PDF poles because we involve the differentiated PDF in the product that we apply the inverse Mellin transform to.} Specifically, we have used $C_{\text{MP}}=2$ in the following. For improved convergence of the integral we also deflect the integration path, on both sides of the real axis, by an angle $\pi/4$ toward the imaginary axis. This allows us to rewrite the form of the inverse Mellin transform \eqref{eq:invmellin} as
\begin{equation}
    f\left(z\right) = \frac{1}{\pi}\int_0^1\frac{dy}{y}\text{Im}\left(e^{i\phi}z^{-N\left(y\right)}F\left(N\left(y\right)\right)\right)\text{,}
\end{equation}
where $N\left(y\right)=C_{\text{MP}}-e^{i\phi}\log{y}$, and $\phi=3\pi/4$. For the integral to converge when $z>1$, the contour must be deflected away from the imaginary axis instead; this amounts to taking $\phi=\pi/4$ above. We discuss the convergence of this integral and the choices made here in more detail in Appendix~\ref{sec:convergence_properties_Mellin}.

\subsection{Comparison to existing calculations}
\label{sec:existing_calculations}
It is worth commenting on how the results of this section compare to those found in existing state-of-the-art calculations for slepton pair production, specifically those implemented in the \textsc{Resummino} tool~\cite{Fiaschi:2023tkq,Fiaschi:2019zgh,Fiaschi:2018xdm,Bozzi:2007tea,Bozzi:2007qr,Bozzi:2006fw}, as well as the leading-power N$^3$LL calculation found in Ref.~\cite{Broggio:2011bd}. 

\textsc{Resummino} offers the calculation of the cross-section for slepton pair production to what the authors term aNNLO+NNLL~\cite{Fiaschi:2019zgh}, where aNNLO stands for ``approximate next-to-next-to-leading order'', and NNLL indicates inclusion also of resummed terms from Ref.~\cite{Vogt:2000ci} that are lower in logarithmic order than the NLL ones, but still at leading power. The code computes full LO and NLO fixed order contributions, allowing for left/right eigenstate mixing between the third generation sfermions. As implemented in the code, in the resummed results the hard function in \eqref{eq:qqres} is modified by adding terms to order $\order{\alpha_s}$ for the NLL result, and terms up to $\order{\alpha_s^2}$ for the NNLL result. The resulting resummed contributions are matched to the fixed order NLO results. It is these higher order effects through the hard function that constitute the terms beyond NLO that the authors denote aNNLO.

By comparison, as our baseline we include all fixed order contributions up to and including NLO, as well as the NLL results, where we have chosen not to include $\order{\alpha_s}$ terms in the hard part of the interaction in \eqref{eq:qqres} for the reasons discussed in Section~\ref{sec:slepton_nlpresult}. Following the notation used in Ref.~\cite{vanBeekveld:2021hhv}, for clarity the resummed results from \textsc{Resummino} are then referred to as NLL' and NNLL' in comparisons below, as such higher-order terms \emph{are} included in their hard functions. Since these terms fully capture the terms beyond NLO in \textsc{Resummino} we will here refer to their results as NLO+NLL' and NLO+NNLL'.

We extend our baseline by including the next-to-leading power corrections discussed in this work that are suppressed by one power in the threshold variable. It has been noted previously~\cite{vanBeekveld:2021hhv} that these power-suppressed terms may be equally significant as the NNLL contributions to the total result, in particular in the regions of phase space that are somewhat further removed from the threshold. This is particularly important for the production of lower-mass particles, since the threshold region is expected to be less dominant in such cases. How these different corrections affect the total cross-section will be discussed  in section \ref{sec:results}. 

To facilitate comparison between our results and those from \textsc{Resummino}, we will also in places include NNLL contributions, and extra $\alpha_s$-suppressed terms in the hard function. The former amounts to adding  a term $\alpha_sg_3\left(\lambda\right)$ to the exponent in \eqref{eq:g1g2h1}~\cite{Catani:2003zt}; this in turn requires the QCD $\beta$-function~\cite{Tarasov:1980au,Larin:1993tp} of \eqref{eq:alphas_beta} and the splitting function of \eqref{eq:splitting_A}~\cite{Vogt:2004mw,Moch:2004pa} to three loops, and the soft wide-angle function of \eqref{eq:resum_D} to two loops~\cite{Vogt:2000ci}. The relevant expressions, including the hard function up to $\order{\alpha_s^2}$, can be found in Appendix D of Ref.~\cite{vanBeekveld:2021hhv}.

There is some degree of overlap between the NLP corrections we include and those of the \emph{collinear-improved} resummation formalism~\cite{Kramer:1996iq,Catani:2001ic}, that is included in \textsc{Resummino} for NLL resummation~\cite{Bozzi:2007qr}. The latter also leads to the resummation of logarithms that are suppressed by a factor of $1/N$ in Mellin space, though they are treated differently, through the use of QCD evolution operators~\cite{Debove:2010kf}. In our results, they arise instead when the kinematically suppressed effects are systematically included in the threshold resummation formalism.

\textsc{Resummino} also handles logarithmic effects from regions of low slepton-pair transverse momentum, by resumming these logarithms to NLL both individually~\cite{Bozzi:2006fw} and jointly with the threshold region discussed here~\cite{Bozzi:2007tea}. Such corrections are not included in our results, as we are mostly focused on the total cross section.

Results to N$^3$LL have also been found using SCET in~\cite{Broggio:2011bd}. These results contain threshold resummed contributions up to one order higher in leading-power logarithms than those available in \textsc{Resummino}, matched to an NLO fixed-order expression. As in \textsc{Resummino}, the hard factor in the resummation formula contains terms up to $\order{\alpha_s^2}$, which are necessary in this case to capture all N$^3$LL logarithms. The NLO fixed-order calculation assumes degenerate squark masses and no squark mixing, but is otherwise identical to the results presented here.

As we noted in Section~\ref{sec:resummation}, we are using threshold resummation mainly as a method to stabilize fixed-order results, and are therefore limiting ourselves to including the logarithmic orders present at NLO. This allows for a more straightforward comparison of the effects of higher power resummation, relative to those of leading-power resummation to higher logarithmic orders. The latter should be expected to provide a more precise estimation of the cross-section for high final-state masses, as these will be dominated by the threshold region. On the other hand, the inclusion of higher-order logarithms than those present in the accompanying fixed-order calculation may actually lead to larger uncertainties as the masses become smaller, due to the additional loop orders needed in the hard function of these expressions. We will return to this point in Section~\ref{sec:scaleunc}.

\section{Numerical results}
\label{sec:results}
In this section, we present results from a numerical evaluation of the above expressions. We focus on smuon pair production $pp\rightarrow \tilde\mu_{L/R}\tilde\mu_{L/R}^*+X$. This process was chosen as it has a simple one-parameter dependence on the mass of the smuon --- excluding the squark and gluino parameters, which enter at NLO order and only minimally affect the results. The impact of the QCD corrections is not anticipated to depend significantly on the final state in slepton pair production. 

\subsection{Details of numerical evaluation}
For comparisons of the results with and without the next-to-leading power (NLP) resummed logarithms included, we label the order of the calculation by the highest logarithmic order that is resummed at each power in the threshold expansion; in this way the results discussed in the previous section are referred to as ``NLO + LP NLL + NLP LL'', while conventional leading-power NLL resummation is denoted by ``NLO + LP NLL''.

Our results are compared with results obtained by running \textsc{Resummino~3.1.2}~\cite{Fiaschi:2023tkq}, for the different combinations of logarithmic contributions available: NLO + NLL', NLO + NNLL', which are pure leading power contributions, as well as the collinear-improved NLO+NLL', which contains terms beyond leading power as discussed in Section~\ref{sec:existing_calculations}. As a reminder, the \textsc{Resummino} NLO + NLL' result is identical to our NLO + LP NLL result except for the addition of extra $\order{\alpha_s}$ terms in the hard ($N$-independent) function in \eqref{eq:qqres}, that are absent in our expression.

Unless otherwise stated, expressions were evaluated using a center-of-mass energy of $\sqrt{s}=13.6$ TeV, and with the \texttt{PDF4LHC21\_40\_pdfas}~\cite{PDF4LHCWorkingGroup:2022cjn} Hessian PDF sets using  \textsc{LHAPDF 6.5.3}~\cite{Buckley:2014ana} to evaluate PDFs and $\alpha_s$.
Passarino--Veltman functions were evaluated using \textsc{LoopTools~2.16}~\cite{Hahn_1999,vanOldenborgh:1989wn}.

Uncertainties due to missing higher-order terms (MHOUs --- missing higher-order uncertainties) are estimated by what is referred to below as scale uncertainty, calculated using the 7-point scale variation. For a central scale $\mu_0$, this means computing the cross-section for each of the following sets of factorization and renormalization scales:
\begin{equation}
    \left(\frac{\mu_F}{\mu_0},\frac{\mu_R}{\mu_0}\right) \in \left\{\left(1,1\right),\left(\frac{1}{2},\frac{1}{2}\right),\left(\frac{1}{2},1\right),\left(1,\frac{1}{2}\right),\left(1,2\right),\left(2,1\right),\left(2,2\right)\right\}\text{,}
\end{equation}
where the range between the largest and smallest estimate is taken as the error band. The central scale is set to $m_{\tilde{\mu}_L}$ for the total cross-section $\sigma$, and to the invariant mass $Q$ for the differential cross-section $d\sigma/dQ$. This estimate of the MHOU is occasionally denoted by $\delta_\mu$ in the following.

It should be noted that this is not a particularly satisfactory way of estimating MHOUs, as it is notoriously unreliable, e.g.\ in cases where different couplings appear at higher perturbative orders, such as the difference between LO and NLO of the present case, and has no proper statistical foundation as an uncertainty. This has been addressed repeatedly in the literature and alternative definitions have been proposed, see, e.g., Refs.~\cite{Cacciari:2011ze,Bonvini:2020xeo,Duhr:2021mfd,Tackmann:2024kci}. Several issues remain, however, such as how to properly account for correlations with MHOUs in the PDF fitting procedure. 
That being said, scale variations remain the current standard method of estimating MHOUs, and are used here for the sake of comparison, in particular to compare the perceived accuracy of existing calculations with the NLP corrections included here.

Uncertainties due to PDFs and $\alpha_s$ are calculated according to the PDF4LHC working group recommendations described in Ref.~\cite{PDF4LHCWorkingGroup:2022cjn}, and combined into what is referred to below as the ``PDF$+\alpha_s$'' uncertainty.

We also include comparisons to the NLO + N$^3$LL results in~\cite{Broggio:2011bd} as discussed previously. These are referred to as ``NLO + LP N$^3$LL,'' for consistency. Since these results are not available in a numerical package, some different conventions have to be used exceptionally for these calculations to make the comparison as close as possible: The central renormalization and factorization scales are set to the invariant mass $Q$ also for the total cross-section; scale uncertainties are estimated by varying both scales simultaneously, not individually as with the 7-point method; and some electroweak parameters are changed, specifically $\alpha$ is fixed by $\alpha\left(m_Z\right)=1/128$, and the weak mixing angle used is $\sin^2\theta_W=0.23143$. These calculations are also made using a different PDF set, \texttt{MSTW2008nnlo}~\cite{Martin:2009iq} for the resummed results and the corresponding NLO set for NLO results. 

The authors of Ref.~\cite{Broggio:2011bd} define two different supersymmetric benchmark points which we use for comparison: The point $P_1$, with 180 GeV sleptons, 600 GeV squarks, and 750 GeV gluinos; and $P_2$, with 360 GeV sleptons, 1200 GeV squarks, and 500 GeV gluinos. In both cases, no squark mixing is included.

It should be noted that the SCET calculations feature three different scales: The factorization scale $\mu_F$, as well as hard and soft scales $\mu_h$ and $\mu_s$. In the results of Ref.~\cite{Broggio:2011bd}, all three of these scales are varied simultaneously to find the scale uncertainty of resummed cross-sections. As a result, the scale uncertainties presented there are not necessarily directly comparable to ours. However, since NLO + NNLL results are available from the same reference,\footnote{These results are only available for the given benchmark points at a few tabulated centre-of-mass energies, so comparisons will be restricted to these.} and we generate our own NLO + LP NNLL results, we can still reliably compare the effects of adding NLP LL or LP N$^3$LL contributions, relative to the respective NLO + NNLL result. 

The results in~\cite{Broggio:2011bd} also include supersymmetric QCD corrections to one-loop in the hard function of the resummed expression, which are not present in our expressions at LP NNLL, and similarly lacking in  \textsc{Resummino}. We have verified numerically that the effect on the total cross-section is minimal.

The datasets used to generate our results in the following figures are available from \textsc{Zenodo}~\cite{klungland_2025_17093804}.

\subsection{Total and differential cross sections}
\begin{figure}[t]
    \centering
    \includegraphics{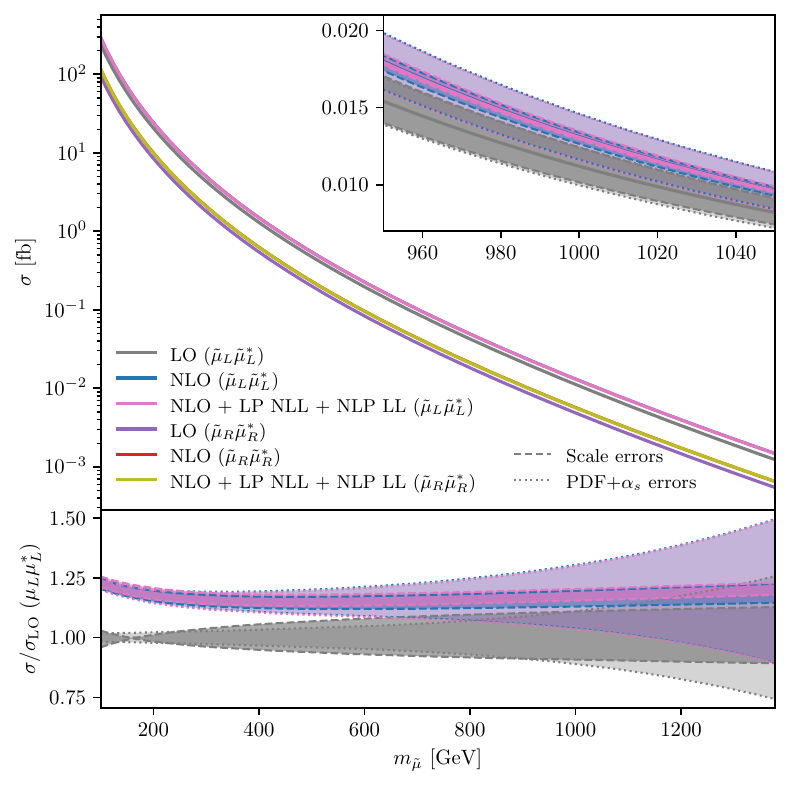}
    \caption{Total cross-sections for smuon pair-production as a function of the smuon mass, at $\sqrt{s}=13.6$ TeV. The main part of the figure shows the central predictions, while the inset (with only $\mu_L\mu_L^*$ production results) also shows the scale and combined PDF and $\alpha_s$ uncertainties as bands for a limited range of masses. Below we show the cross-sections and uncertainties relative to the LO central prediction, again for $\mu_L\mu_L^*$. Here the central predictions are omitted, to prevent clutter.}
    \label{fig:total_cs_LHC}
\end{figure}

\begin{figure}[t]
    \centering
    \includegraphics{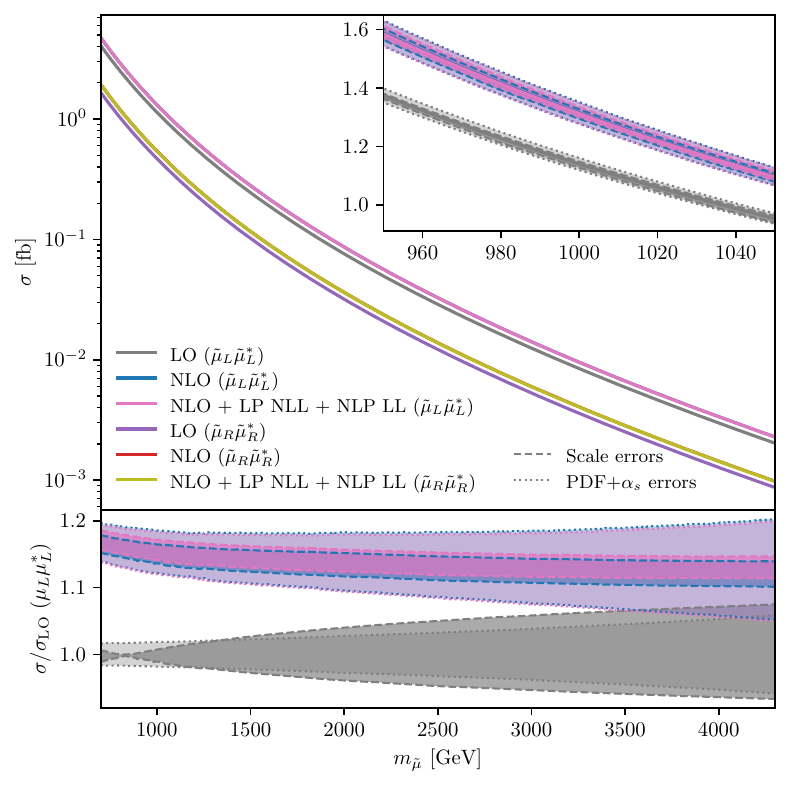}
    \caption{Similar to Figure~\ref{fig:total_cs_LHC}, but with $\sqrt{s}=85$ TeV.}
    \label{fig:total_cs_FCChh}
\end{figure}

The total cross-section as a function of the smuon mass is shown for the production of left- and right-handed smuons at $\sqrt{s}=13.6$ TeV, in Figure~\ref{fig:total_cs_LHC}. Here, the squark masses are set to $m_{\tilde{q}}=2$ TeV, and the gluino masses to $m_{\tilde{g}}=2.5$ TeV, while the $\tilde{b}$ and $\tilde{t}$ mixing matrices are assumed diagonal, i.e.\ $\mathbf{b}_{AB}=\mathbf{t}_{AB}=\delta_{AB}$ in \eqref{eq:slepton_mixing_matrix}, although the effect of squark mixing is numerically completely negligible.
We show both the fixed order results and the final resummed results up to next-to-leading power at leading logarithmic accuracy. Similar results for the FCC-hh centre-of-mass energy $\sqrt{s} = 85$ TeV are shown in Figure~\ref{fig:total_cs_FCChh}.

\begin{figure}[t]
    \centering
    \includegraphics{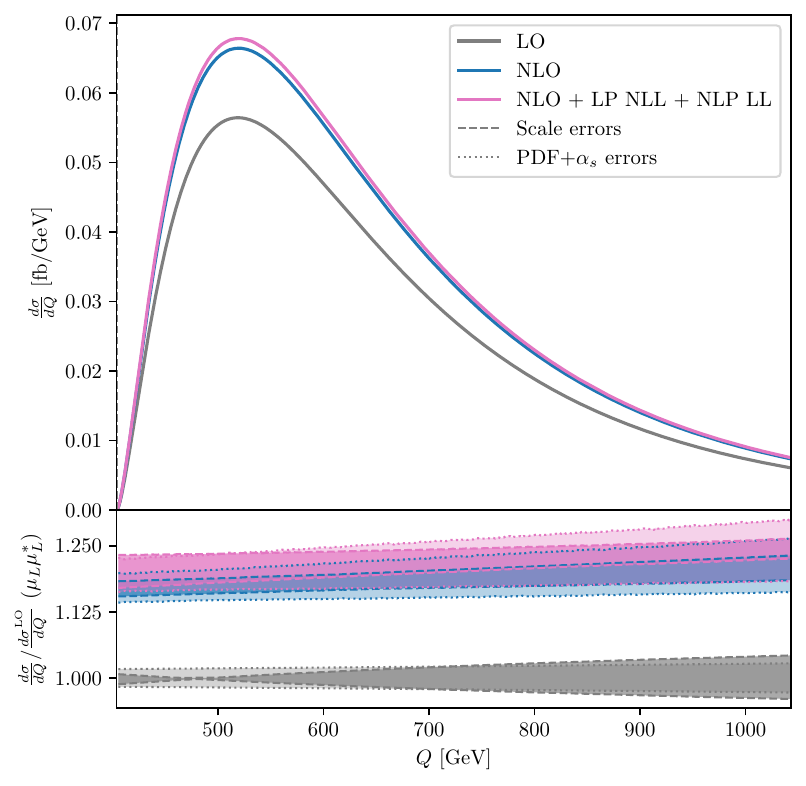}
    \caption{Differential cross-section for $\mu_L\mu_L^*$ production in the smuon-pair invariant mass $Q$, for the SPS1a benchmark point at $\sqrt{s}=13.6$ TeV. The top part of the figure shows the central predictions, while below we show the cross-sections and uncertainties relative to the LO central prediction.
    }
    \label{fig:diffcs_sps1a}
\end{figure}

\begin{figure}[ht]
    \centering
    \includegraphics{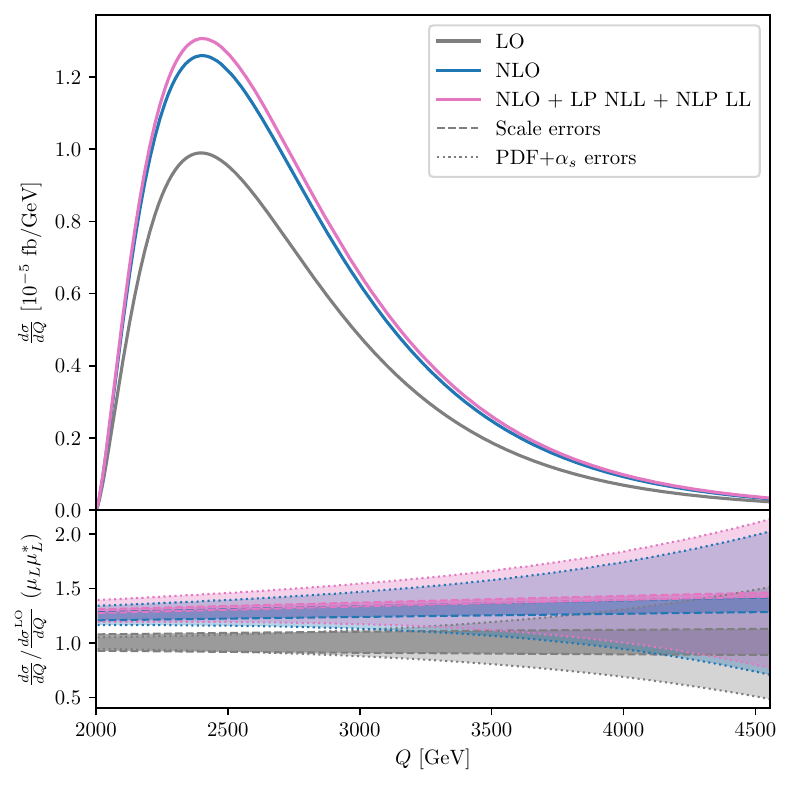}
    \caption{Similar to Figure~\ref{fig:diffcs_sps1a}, but for scenario with $m_{\tilde{\mu}_L}=1000$~GeV.}
    \label{fig:diffcs_resumminopaper}
\end{figure}

Plots of the differential cross-section (at $\sqrt{s}=13.6$ TeV, for left-handed smuon production) in terms of the invariant mass can be found in Figure~\ref{fig:diffcs_sps1a} for the SPS1a benchmark point~\cite{Allanach:2002nj} which features smuons with masses $m_{\tilde{\mu}_L}=202$ GeV and $m_{\tilde{\mu}_R}=143$ GeV, and in Figure~\ref{fig:diffcs_resumminopaper} with $m_{\tilde{\mu}}=1$ TeV and the squark and gluino masses set as above. Uncertainties from both scale variation and PDF$+\alpha_s$ are included in Figs.~\ref{fig:total_cs_LHC}--\ref{fig:diffcs_resumminopaper}; however, recall that the scale error is computed in the same way for both calculations, but the central scale is different for total and differential cross-sections. 

These results demonstrate the significant reduction in scale error when including resummed contributions already observed in the literature~\cite{Bozzi:2007qr}. This effect gets more pronounced when the produced particles are heavier, as anticipated since the threshold region is more significant kinematically in this case. As we can see in Figure~\ref{fig:total_cs_LHC}, at $\sqrt{s}=13.6$ TeV the scale error is dominated by the PDF error as the smuon masses exceed 1 TeV, due to kinematical constraints requiring PDFs to be sampled in higher-uncertainty regions, lessening the impact of reducing the scale error on the total theoretical error. For the $\sqrt{s}=85$ TeV case the scale error is of the same order of magnitude as the PDF error even at smuon masses of multiple TeV, indicating the importance of reducing the scale error for slepton searches at future colliders.


To compare with existing leading-power resummed results to N$^3$LL from SCET, we also list our total cross-section results at $\sqrt{s}=7$ and 14 TeV in Table~\ref{tab:N3LLcomp}, for the parameter points $P_1$ and $P_2$ defined in Section~\ref{sec:existing_calculations}. We recall that these calculations were done with different conventions than for the other results presented here, notably with the central scales at $\mu_0=Q$ and with the scale uncertainty estimated by simultaneous variation of both the renormalization and factorization scales. We include our results at NLO, NLO + LP NNLL, and NLO + LP NLL + NLP LL; these are compared to results obtained from Ref.~\cite{Broggio:2011bd} to NLO, NLO + LP NNLL, and NLO + LP N$^3$LL. 

\begin{table}[ht]
    \centering
    \begin{tabular}{l|P{0.2\textwidth}|P{0.2\textwidth}|P{0.2\textwidth}}
        \hline
         & $\sqrt{s}=7$ TeV, point $P_1$ & $\sqrt{s}=14$ TeV, point $P_1$ & $\sqrt{s}=14$ TeV, point $P_2$\\\hline
         Our results & & & \\
         $\sigma_{\text{NLO}}$ & $10.56^{+0.24}_{-0.21}$ & $36.68^{+0.42}_{-0.34}$ & $2.46^{+0.04}_{-0.05}$\\
         $\sigma_{\text{NLO + LP NNLL}}$ & $10.76^{+0.15}_{-0.14}$ & $37.57^{+0.73}_{-0.66}$ & $2.50^{+0.02}_{-0.03}$\\
         $\sigma_{\text{NLO + LP NLL + NLP LL}}$ & $10.78^{+0.13}_{-0.01}$ & $37.65^{+0.69}_{-0.22}$ & $2.50^{+0.03}_{-0.01}$\\ \hline
         From \cite{Broggio:2011bd} & & & \\
         {$\sigma_{\text{NLO}}$} & {$10.56^{+0.24}_{-0.22}$} & {$36.65^{+0.45}_{-0.35}$} & {$2.45^{+0.05}_{-0.05}$}\\
         {$\sigma_{\text{NLO + LP NNLL}}$} & {$10.63^{+0.13}_{-0.17}$} & {$37.16^{+0.36}_{-0.46}$} & {$2.47^{+0.03}_{-0.03}$}\\
         {$\sigma_{\text{NLO + LP N$^3$LL}}$} & {$10.81^{+0.10}_{-0.08}$} & {$37.80^{+0.25}_{-0.12}$} & {$2.51^{+0.02}_{-0.02}$}\\\hline
    \end{tabular}
    \caption{Total cross-section for $\tilde{\mu}_L\tilde{\mu}_L^*$ production in fb, with associated scale uncertainties, for varying center-of-mass energies and different parameter points as defined in Section~\ref{sec:existing_calculations}.}
    \label{tab:N3LLcomp}
\end{table}

The NLO results are included to verify the compatibility of the fixed-order calculations, as they indeed do. The leading-power NNLL results from both sources are included as a frame of reference when looking at the scale uncertainties of the resummed results, since the results in SCET depend on the extra soft scale $\mu_s$. We see that also the central values here differ somewhat despite the calculations being to the same logarithmic order. This is not too surprising due to the different formalisms used in the two calculations, including the soft scale present in SCET.

Both the LP N$^3$LL and our LP NLL + NLP LL results exhibit a fairly significant reduction in scale uncertainty relative to the corresponding LP NNLL result for all the test points. Quantitatively this reduction is quite similar for the two approaches, showing that they are both valuable approaches to reducing the scale error. For both test points the masses of the produced sleptons are relatively light, thus the effect of NLP resummation, which also captures kinematical regions somewhat further removed from the threshold, might be relatively more significant here than for large masses. This is further born out by the reduction being smaller for the highest mass point $P_2$. As expected, for all the test points the higher logarithmic corrections bring the central values into better agreement.

\subsection{Scale dependence}
\begin{figure}[ht]
    \centering
    \includegraphics{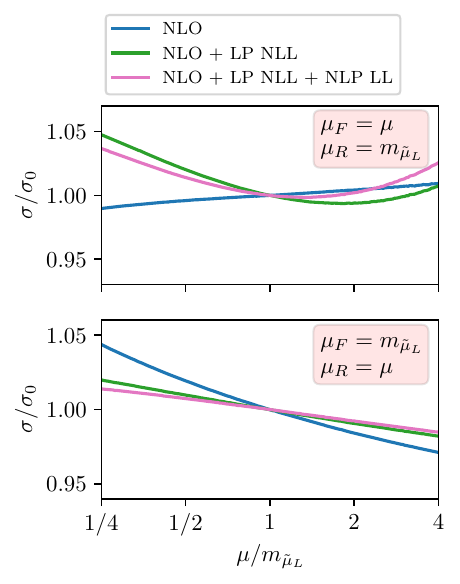}
    \includegraphics{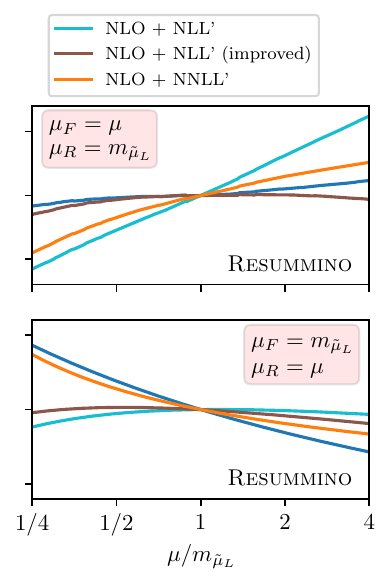}
    \caption{Cross-sections for $\tilde{\mu}_L\tilde{\mu}_L^*$ production with $m_{\tilde{\mu}_L}=120$ GeV, at various orders, as a function of renormalization and factorization scales, normalized by their values at the central scale $\mu_{F/R}=m_{\tilde{\mu}_L}$ (``$\sigma_0$''). Left: Our results. Right: Obtained from \textsc{Resummino}.}
    \label{fig:scale_var_120}
\end{figure}

\begin{figure}[ht]
    \centering
    \includegraphics{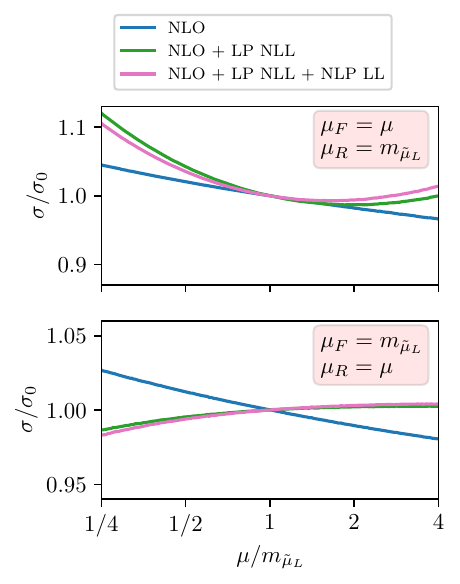}
    \includegraphics{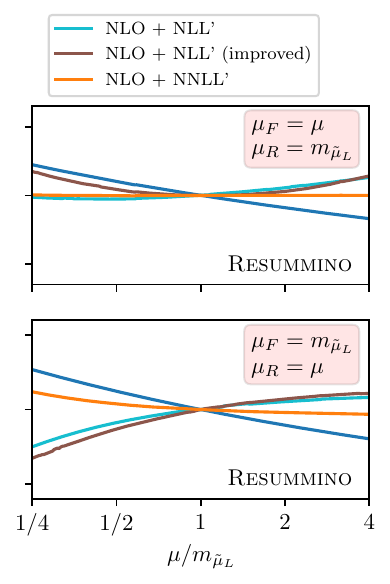}
    \caption{Same as Figure~\ref{fig:scale_var_120}, but with $m_{\tilde{\mu}_L}=1000$ GeV.}
    \label{fig:scale_var_1000}
\end{figure}

Individual breakdowns of the renormalization and factorization scale dependence of the total cross-section, for fixed values of the mass, are shown in Figure~\ref{fig:scale_var_120} (for $m_{\tilde{\mu}_L}=120$ GeV) and Figure~\ref{fig:scale_var_1000} ($m_{\tilde{\mu}_L}=1000$ GeV). We include the fixed next-to-leading order (NLO), the leading-power next-to-leading logarithmic order (LP NLL), and the next-to-leading power results (NLP LL) presented in this paper.
These results are compared to similar results from \textsc{Resummino}. In the lighter-smuon case we have set gluino and squark masses to 2~TeV and 1~TeV, respectively, while for the heavier case these are all set to 1.3~TeV. These numerical choices follow what was done in Ref.~\cite{Fiaschi:2019zgh} to ease comparison.  We have verified numerically that the relative difference between our NLO calculation and that of \textsc{Resummino} is $\order{10^{-3}}$ and consistent with the numerical uncertainty of the PDF integral.

The renormalization dependence of our results are largely what one would naively expect, with a sizeable decrease in dependence from NLO to NLO+NLL, and a smaller decrease as NLP corrections are included. 

The factorization scale dependence is less straightforward to understand, as this depends on the interplay between scale dependencies of the partonic cross-section and the PDFs.\footnote{For consistency, we should ideally be using PDF sets that incorporate threshold resummation effects matching those in the partonic cross sections. However, even though such sets have recently been developed~\cite{Bonvini:2015ira}, and used for resummed slepton pair production cross-sections~\cite{Fiaschi:2018xdm}, a lack of available resummed calculations means that a number of processes must currently be left out of the fit~\cite{Bonvini:2015ira}. This leads to a significantly smaller dataset, and a larger PDF uncertainty. Thus the overall uncertainty would be larger, even with a potentially smaller factorization scale dependence.} We see a larger factorization scale dependence when including the resummed contributions, though we still observe a slightly smaller scale dependence in the NLP result compared to the LP one. Qualitatively similar results are found with \textsc{Resummino}. 

Functionally, the increased $\mu_F$ dependence of the resummed results compared to the fixed-order one stems from the presence of the factor $\log{\frac{\mu_F^2}{Q^2}}$ in the NLL exponent, see Eq.~\eqref{eq:g2}. This could potentially be compensated for by the inclusion of $\order{\alpha_s}$ terms in the hard function of the resummation formula, as is done in \textsc{Resummino} for the NLO+NLL' result. However, looking at the $\mu_F$ dependence of the NLO+NLL' result at both values of the smuon mass, this works at high smuon masses as pointed out in \cite{Fiaschi:2019zgh}, but does not seem to give a significantly less $\mu_F$-dependent result at low masses. The same is true even for the NLO+NNLL' result, which also has a fairly significant factorization scale dependence for $m_{\tilde{\mu}_L}=120$ GeV. The low scale dependence for large masses, as is seen in Figure~\ref{fig:scale_var_1000} for both the NLO+NLL' and NLO+NNLL' results, might also lead to underestimation of the uncertainty from higher orders. We will revisit this point in the next section.

It is also worth looking at the scale dependence of the improved resummation result obtained from \textsc{Resummino}. For the relatively light final-state masses in Figure~\ref{fig:scale_var_120}, it is remarkably flat as a function of both scales, as pointed out in the original paper~\cite{Bozzi:2007qr}. For higher masses, though, as in Figure~\ref{fig:scale_var_1000}, we find that the dependence is more similar to the other orders, suggesting that the benign lower-mass behavior may be a coincidence in the functional form of the cross-section.

\begin{figure}[ht]
    \centering
    \includegraphics{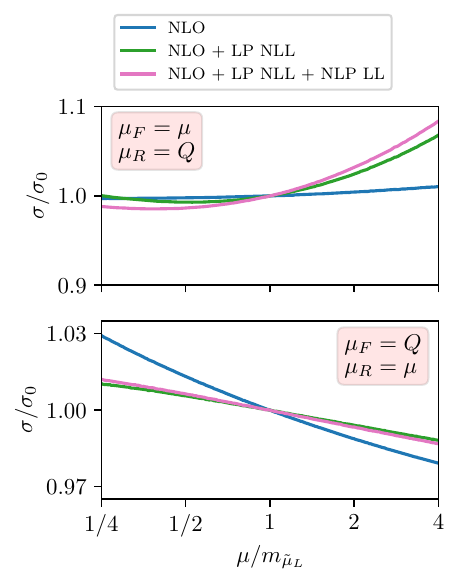}
    \includegraphics{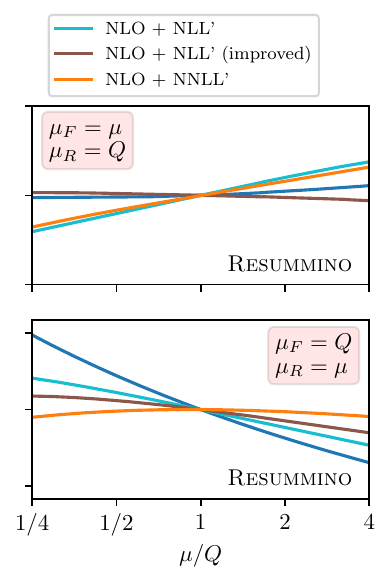}
    \caption{Similar to Figure~\ref{fig:scale_var_120}, but for the differential cross-section at $Q=500$ GeV, and with the central scale at $\mu_{F/R}=Q$.}
    \label{fig:scale_var_diff_sps1a}
\end{figure}

\begin{figure}[ht]
    \centering
    \includegraphics{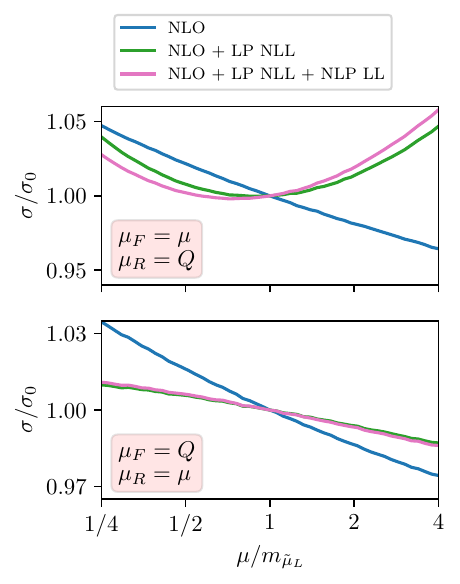}
    \includegraphics{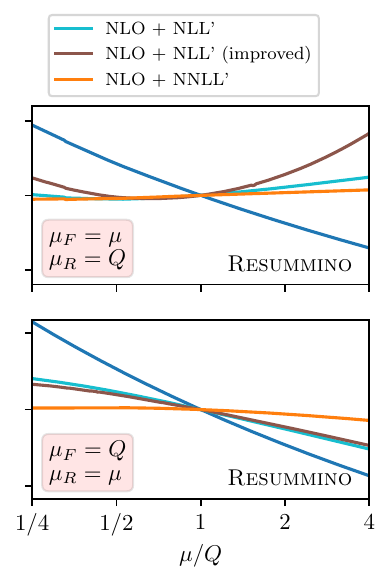}
    \caption{Same as Figure~\ref{fig:scale_var_diff_sps1a}, but with $m_{\tilde{\mu}_L}=1000$ GeV and $Q=2500$ GeV.}
    \label{fig:scale_var_diff_resumminopaper}
\end{figure}

Turning to the differential cross-section, the renormalization and factorization scale dependence of the various results is shown in Figure~\ref{fig:scale_var_diff_sps1a} for the SPS1a benchmark point with $Q=500$ GeV, and in Figure~\ref{fig:scale_var_diff_resumminopaper} with $m_{\tilde{\mu}_L}=1$ TeV and $Q=2.5$ TeV. Note that the scales are varied around $Q$ rather than the smuon mass in these calculations. The dependence is mostly similar to that of the total cross-section, but unlike for the total cross-section, the NLP result actually has a somewhat larger dependence than the LP one. The same was observed for the differential cross-section in Ref.~\cite{vanBeekveld:2021hhv}, where the authors suggested that the scale dependence would be expected to decrease once the NLP results are available at the same logarithmic order as the LP results.


\subsection{Scale uncertainty}
\label{sec:scaleunc}
\begin{figure}[ht]
    \includegraphics[scale=1.09]{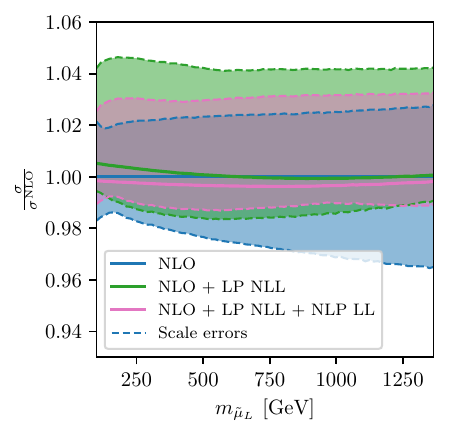}
    \includegraphics[scale=1.09]{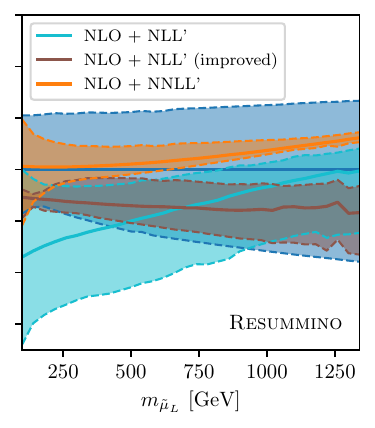}
    \caption{$K$-factor relative to the NLO cross-section of the resummed results at various orders, along with the 7-point scale uncertainty. Left: The results of this paper. Right: Results obtained from \textsc{Resummino}.}
    \label{fig:ratio_scaleunc}
\end{figure}

\begin{figure}[ht]
    \includegraphics[scale=1.09]{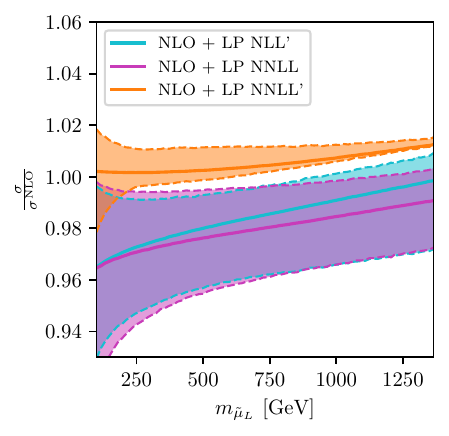}
    \includegraphics[scale=1.09]{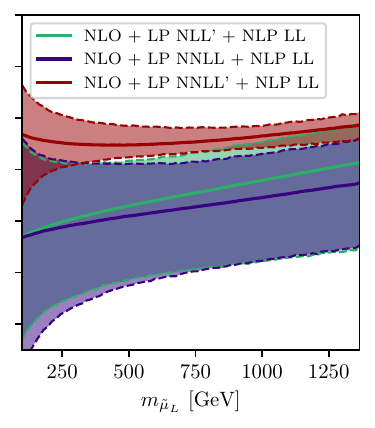}
    \caption{Our results for cross-sections before (left) and after (right) adding NLP corrections to various LP results, that extends our LP baseline of NLO+NLL to include higher logarithms and corrections to the hard function. All results are relative to the NLO result and include the 7-point scale uncertainty. Our NLO+LP NLL' (cyan) and NLO+LP NNLL' (orange) results in the left-hand plot are identical to the same \textsc{Resummino} results in Fig.~\ref{fig:ratio_scaleunc}.}
    \label{fig:LP_NLP_combinations}
\end{figure}

On the left-hand side of Figure~\ref{fig:ratio_scaleunc} we show the impact of the various resummed contributions on the $\tilde{\mu}_L\tilde{\mu}_L^*$ cross-section and its scale uncertainty, as a function of the smuon mass $m_{\tilde{\mu}_L}$.\footnote{All other parameters are, for our amusement, set to their values at the SPS1a benchmark point~\cite{Allanach:2002nj}. However, the impact is negligible.}   
Some further results, adding NLP corrections to higher resummed orders, are shown in Figure~\ref{fig:LP_NLP_combinations}. The calculations with leading and next-to-leading power logarithms, normalized to the fixed order NLO result, are shown accompanied by their 7-point scale variation uncertainties. On the right-hand side of Figure~\ref{fig:ratio_scaleunc} we show corresponding results from \textsc{Resummino}. Note that each set of results is normalized to its respective NLO calculation.

We observe a small decrease in the calculated scale uncertainty in our results as the order of the calculation is increased---both as we go from fixed-order to resummed results, and as NLP corrections are included---as one would anticipate. In particular, the size of the uncertainty remains relatively constant as a function of the smuon mass for the resummed results, at the level of $(+3\%, -1\%)$ when including the NLP contributions, whereas it grows for the fixed-order one. Thus the improvement when including the resummed terms is more significant in the high mass region, as anticipated at the end of Section~\ref{sec:fixed_threshold}. The shift in the central cross-section prediction is relatively minor, though we note that the size of the shift resulting from the inclusion of NLP corrections is fairly similar to the difference between the NLO and the leading-power NLO+NLL results, and consistently lowers the cross section.

Comparing to \textsc{Resummino}, we notice that the computed scale uncertainty for large masses is smaller than in our results, and that both of the ``unimproved'' results, i.e.\ NLO+NLL' and NLO+NNLL', show a significantly decreasing uncertainty for larger masses. The same pattern can also be observed when NLP terms are added to these, as shown in Figure~\ref{fig:LP_NLP_combinations}.
This is particularly apparent for the NLO+NNLL' result from \textsc{Resummino} in Figure~\ref{fig:ratio_scaleunc}, where the relative scale uncertainty is well below $10^{-2}$ for smuon masses above 1 TeV. As can be seen from the high-mass region of the left-hand plot in Figure~\ref{fig:ratio_scaleunc}, the negative shift in the central cross-section prediction as NLP corrections are taken into account---which are \emph{not} included in the NLO+NNLL' result---is larger than the lower scale uncertainty of the \textsc{Resummino} prediction. This can also be observed in Figure~\ref{fig:LP_NLP_combinations}, where the scale uncertainty for the NLO+NNLL' result increases significantly when the NLP contribution is added (right-hand plot).
This indicates that the scale uncertainty may be underestimated in the \textsc{Resummino} NLO+NNLL' result, which is the current state-of-the-art for numerical slepton cross section predictions.

\begin{figure}[ht]
    \centering
    \includegraphics[scale=0.95]{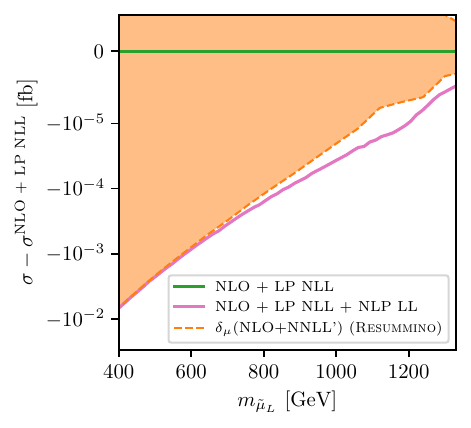}
    \includegraphics[scale=0.95]{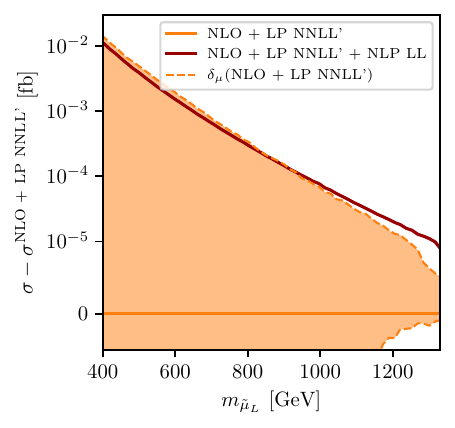}
    \caption{Shift in the cross-section when adding NLP corrections to leading-power NLL (left) and NNLL' (right) resummed results. Overlaid in orange is the scale uncertainty of the leading-power NLO+NNLL' result obtained from \textsc{Resummino} (left) and our calculation (right).}
    \label{fig:comp_NLP_NNLL}
\end{figure}

This is made clearer in the left part of Figure~\ref{fig:comp_NLP_NNLL}, which compares the (absolute) shift induced by the NLP contributions on our leading power NLL result, to the NLO+NNLL' scale uncertainty for large masses, indicating that the latter generally fails to predict the former in this mass range, with differences between the scale uncertainty and the shift due to the NLP contribution exceeding a factor of two. Note that since the leading-power logarithmic precision is different, there is no reason to expect our full LL NLP + NLL LP result to lie inside the error band of the full \textsc{Resummino} NLO+NNLL' result. However, given that the latter result does not contain NLP leading logarithmic corrections, the \emph{shift} in the cross-section prediction, which is what Figure~\ref{fig:comp_NLP_NNLL} investigates, should still have approximately the same size. This is further evidenced by the right part of Figure~\ref{fig:comp_NLP_NNLL}, which shows the shift by adding NLP contributions to our NLO + LP NNLL' calculations. Again the NLP corrections are significantly larger than the scale error predicted from the LP NNLL' result at high masses. Note here that the scales here are logarithmic.

Also worth noting in Figure~\ref{fig:LP_NLP_combinations}, in particular for the results with LP NNLL' corrections, is the scale uncertainty increasing significantly for lower masses. These results contain terms in the hard function of $\order{\alpha_s^2}$, from purely virtual corrections, which can be numerically significant even away from the threshold limit when the soft factor of the cross-section is close to 1 (see, e.g., \eqref{eq:NLPres}). Since the fixed-order calculation stops at $\order{\alpha_s}$, these terms will then be the only $\order{\alpha_s^2}$ terms that contribute away from the threshold limit, even though they do not necessarily approximate the overall $\order{\alpha_s^2}$ contributions to the cross-section well (there may be significant contribution from radiation of relatively high-energy color-charged particles, for example). Since these terms contain logarithms of the scale (see, e.g., Refs.~\cite{Kidonakis:2003tx,Kidonakis:2007ww} for the explicit expressions), in some places even squared, it follows that the result may be significantly scale dependent when potentially balancing $\order{\alpha_s^2}$ terms are absent. The same issue might also arise in the NLO+N$^3$LL cross-section for low masses, since two-loop terms in the hard function are required for these calculations, and might in general be a problem if the resummed orders far exceed the logarithmic orders present in the fixed-order calculation. Using resummation mainly as a way of reducing the uncertainty of a perturbative calculation and resumming only the logarithms present in that calculation might then lead to more stable results also for lower masses.

\begin{figure}[ht]
    \centering
    \includegraphics{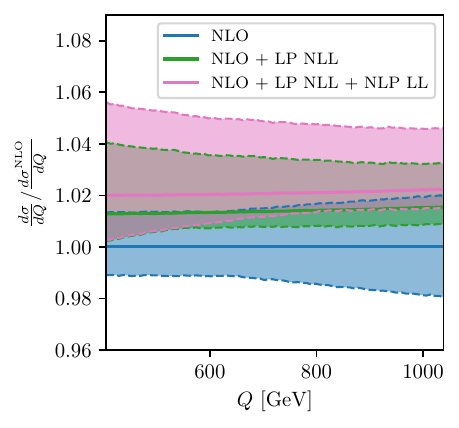}
    \includegraphics{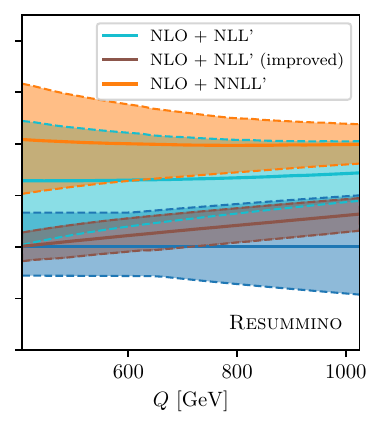}
    \caption{Similar to Figure~\ref{fig:ratio_scaleunc}, but with the differential cross-section as a function of the smuon-pair invariant mass $Q$, for the SPS1a benchmark point.}
    \label{fig:diff_resummino_comp_sps1a}
\end{figure}

\begin{figure}[ht]
    \centering
    \includegraphics{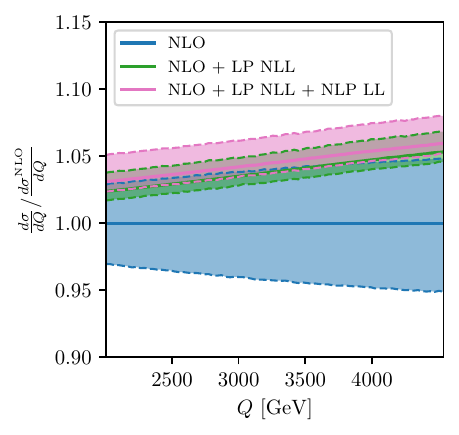}
    \includegraphics{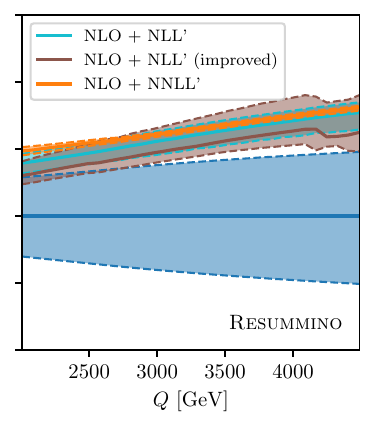}
    \caption{Same as Figure~\ref{fig:diff_resummino_comp_sps1a}, but with $m_{\tilde{\mu}_L}=1000$ GeV.}
    \label{fig:diff_resummino_comp_resumminopaper}
\end{figure}

Similar comparisons of the $K$-factors and scale uncertainties for the differential cross-section, are shown in Figure~\ref{fig:diff_resummino_comp_sps1a} for the SPS1a point, and in Figure~\ref{fig:diff_resummino_comp_resumminopaper} for $1$ TeV smuons. We notice that the scale uncertainty is again similar for the full result and the leading-power NLL result, however, as noted previously, it actually increases slightly for the full result when the scales are varied around the central scale $Q$. For the improved NLO+NLL' calculation we observe some numerical instability in the \textsc{Resummino} results above 4 TeV for both the central value and the scale uncertainty which is visible in Figure~\ref{fig:diff_resummino_comp_resumminopaper} (right). 


\begin{figure}[ht]
    \centering
    \includegraphics[scale=0.95]{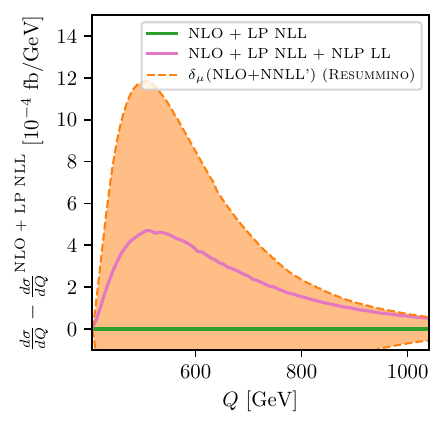}
    \includegraphics[scale=0.95]{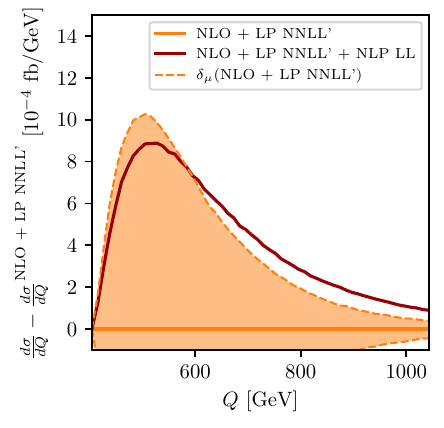}
    \caption{Similar comparison as in Figure~\ref{fig:comp_NLP_NNLL}, but using the differential cross-section $d\sigma/dQ$ for the SPS1a benchmark point.}
    \label{fig:comp_NLP_NNLL_diff_sps1a}
\end{figure}

\begin{figure}[ht]
    \centering
    \includegraphics[scale=0.95]{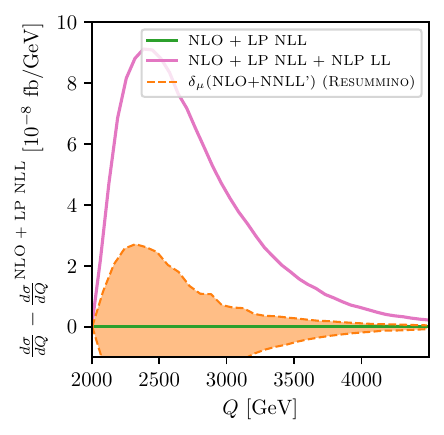}
    \includegraphics[scale=0.95]{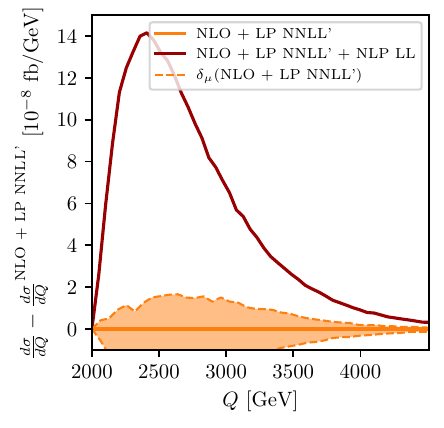}
    \caption{Same as Figure~\ref{fig:comp_NLP_NNLL_diff_sps1a}, but for $m_{\tilde{\mu}_L}=1$ TeV.}
    \label{fig:comp_NLP_NNLL_diff_resumminopaper}
\end{figure}

As for the total cross-section case, we see that the NLO+NNLL' \textsc{Resummino} result has a strikingly small scale uncertainty for high masses, significantly smaller than the impact of the NLP contribution. This is highlighted further in Figure~\ref{fig:comp_NLP_NNLL_diff_resumminopaper}, which illustrates how the estimated MHOU is consistently smaller than the NLP shift for this mass. For completeness, we also include a similar plot for the SPS1a point in Figure~\ref{fig:comp_NLP_NNLL_diff_sps1a}, where the NLO+NNLL' uncertainty band covers the NLP shift more successfully.


The interpretation of these results is not immediately obvious. They seem to suggest that the existing NLO+NNLL' result somehow ``artificially'' has a scale dependence that is unrealistically low---either due to some issue with the construction of the result, e.g.\ the addition of $\order{\alpha_s^2}$ terms to the hard function in the resummation formula, or the lack of a full fixed-order NNLO cross-section to match against. However, it might also simply be an example of the shortcoming of the conventional approach of estimating MHOUs by scale variation, as noted above.

Regardless of the interpretation, however, this does show that the addition of NLP logarithms has an important effect on the predicted cross-section for slepton pair production. Not so much on the central value---though neither does the addition of LP NLL contributions---but on the assumed uncertainty.
For this reason, the results contained in this paper are an important complement to the standard approach of increasing logarithmic precision, which may not capture all relevant effects.

\subsection{Impact on PDF uncertainty}
In addition to altering the scale uncertainty of the predicted cross-section, the addition of NLP terms can also, at least in principle, affect the PDF uncertainty of the result. While the leading-power resummation expressions only apply to $q\overline{q}$ initial states, at NLP there is an additional contribution from $qg$ and $\overline{q}g$ states. Thus this part of the cross-section calculation will be accompanied by gluon PDFs, whereas the LP resummation contribution only involves $q$ and $\overline{q}$ distributions. Since the gluon PDFs, and their uncertainties, generally have different behaviors~\cite{PDF4LHCWorkingGroup:2022cjn}, this might potentially affect the PDF uncertainty of the overall result. 

\begin{figure}[ht]
    \centering
    \includegraphics{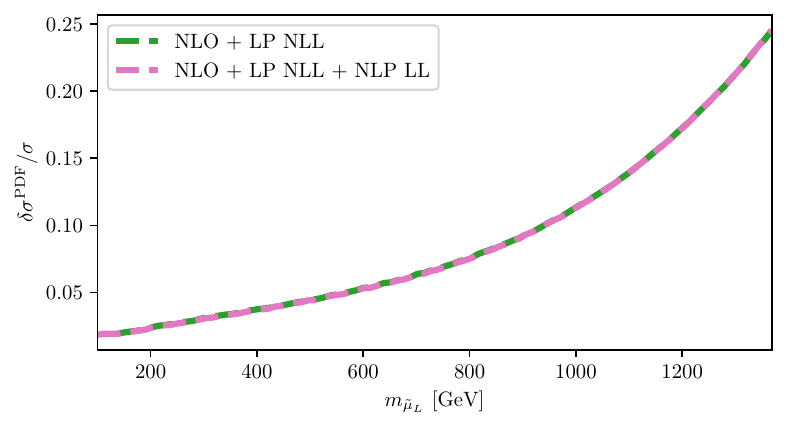}
    \caption{Comparison of the relative PDF uncertainty for the total $\tilde{\mu}_L\tilde{\mu}_L^*$ production cross-section with and without NLP corrections, as a function of the smuon mass.}
    \label{fig:PDFerr}
\end{figure}

A comparison of the relative PDF uncertainties of the LP and NLP results\footnote{Specifically, the NLO + LP NLL and the NLO + LP NLL + NLP LL results, respectively.} in $\tilde{\mu}_L\tilde{\mu}_L^*$ production at $\sqrt{s}=13.6$~TeV, as a function of the smuon mass, is shown in Figure~\ref{fig:PDFerr}. We observe no appreciable change in the uncertainty as the NLP contributions are taken into account. As explained above, any changes to the PDF uncertainty would be associated only with the resummation correction to the cross-section, which as seen in Figure~\ref{fig:ratio_scaleunc} is fairly small compared to the NLO result. Furthermore, the numerical difference between the total cross-section with and without NLP corrections is also relatively small, making any potential impact on the PDF uncertainty introduced at this order small. 

\section{Conclusion}
\label{sec:conclusion}

In this paper we have evaluated the next-to-leading power leading-logarithmic QCD contributions to the inclusive pair production of sleptons, the scalar supersymmetric partners of leptons, at hadron colliders. We have also sought to provide a short pedagogical introduction to threshold resummation at both leading and next-to-leading power, and the practical aspects of numerically evaluating the resummed contributions. To our knowledge, this is the first application of next-to-leading power techniques to the calculation of supersymmetric production cross sections, and BSM cross sections more generally.

Power-suppressed logarithmic corrections have recently been shown to compete with leading-power logarithmic corrections at next-to-leading logarithmic order for SM processes~\cite{vanBeekveld:2021hhv}. We find a similar picture for slepton pair production. The next-to-leading power leading logarithmic corrections are in general of the same size as the leading-power next-to-leading logarithmic corrections. The inclusion of next-to-leading power corrections also tends to reduce the scale dependence of the cross sections, and stabilize the scale dependence as a function of the slepton mass. Comparing to existing results in the literature for next-to-next-to-next-to-leading logarithms at leading-power calculated using SCET, we found that the reduction in scale dependence is similar between the two corrections. The potential effect of the next-to-leading power corrections on PDF errors was found to be completely negligible.

The leading-power result calculated using the \textsc{Resummino} code that we mainly compare to in this paper represents the current state-of-the-art numerical cross-section calculation for slepton pair production, used for example by the LHC SUSY Cross Section Working Group.\footnote{\url{https://twiki.cern.ch/twiki/bin/view/LHCPhysics/SUSYCrossSections}} We find that this result underestimates the scale uncertainty for high slepton masses as the next-to-leading power contributions we find are in general larger than the predicted scale uncertainty. Thus, conventional leading-power resummation may leave out important contributions, even if results are superficially more precise.

One might object that this discussion is somewhat pedantic as the missing higher order uncertainty (MHOU), estimated by the scale uncertainty, is only a sub-leading contribution to the current overall theoretical uncertainty on slepton pair production for large slepton masses at the LHC, compared to the much larger PDF uncertainty. 
Thus, even if the MHOU is somewhat underestimated, this will only have a small effect on the total theoretical uncertainty for these masses. 

We would argue, however, that given the fundamentally different origin of these two sources of uncertainty, it is important to keep both as accurate as possible. If the scale uncertainty is to represent the MHOU, it needs to be an accurate representation of the size of higher-order contributions, even if the impact on the reported total uncertainty is small. The uncertainty on the PDFs at large $x$ will also decrease in the coming years, as more experimental data become available, e.g.\ with higher-luminosity LHC runs, and with possible future lepton--hadron colliders.

Furthermore, should higher-energy hadron colliders such as the FCC-hh be built in the future, these would allow for the production of heavier sleptons at much lower momentum fractions, where the PDF uncertainties are significantly smaller. In this paper we have calculated the cross section and relevant theoretical uncertainties for slepton pair production at a $\sqrt{s}=85$ TeV FCC-hh machine. Here, the MHOU dominates the uncertainty for masses just beyond the reach of the LHC, and accurately determining its size will be critical for the correct interpretation of searches, or, in the case of discoveries, precision measurements.

All of this makes the calculation of power-suppressed logarithmic corrections a worthwhile alternative to the ``standard'' approach of increasing the logarithmic orders. We plan to include the results shown here in a future tool for the calculation of BSM cross sections~\cite{smoking}.

\appendix

\section{Full expressions for cross-section results}
\label{sec:expressions}

For reference, we state here the complete expressions needed for the evaluation of the differential cross-section for slepton pair production to fixed NLO, plus leading power NLL and next-to-leading power LL resummed contributions.

The result is given in terms of the partonic coefficient function $w_{ij}$ in Eq.~(\ref{eq:cs_PDFintegral}), which can be decomposed as $w_{ij}=w_{ij}^{\text{LO}}+w_{ij}^{\text{NLO}}+w_{ij}^{\text{match}}$, where $w_{ij}^{\text{match}}$ is the resummed contribution matched to $\order{\alpha_s}$ of the fixed order contributions, see Section~\ref{sec:matching} for details. The various components are given by:\footnote{We suppress some arguments for readability, including the scale dependence of the strong coupling $\alpha_s\equiv\alpha_s\left(\mu_R\right)$.}
\begin{align}
    w_{q_i\bar{q}_j}^{\text{LO}} &= F^{\tilde{\ell}^k_A\tilde{\ell}'^k_B}_{q_iq_j}\left(Q^2\right)\delta\left(1-z\right)\text{,}\\
    w_{q_ig}^{\text{LO}} &=w_{\bar{q}_ig}^{\text{LO}}= 0\text{,}\\
    \frac{w_{q_i\bar{q}_j}^{\text{NLO}}}{C_F\frac{\alpha_s}{\pi}} &= F^{\tilde{\ell}^k_A\tilde{\ell}'^k_B}_{q_iq_j}\left(Q^2\right) \left\{-\frac{\left(1+z^2\right)\log{z}}{1-z}+2\left(1+z^2\right)\left(\frac{\log{\left(1-z\right)}}{1-z}\right)_++\delta\left(1-z\right)\left(\frac{\pi^3}{3}-4\right)\right.\notag\\
    &\qquad\left.+\log{\frac{\mu_F^2}{Q^2}}\left(-\frac{1+z^2}{\left(1-z\right)_+}-\frac{3}{2}\delta\left(1-z\right)\right)\vphantom{\frac{\left(1+z^2\right)\log{z}}{1-z}}\right\}\notag\\ &\qquad+\sum_{C,D=1}^2F^{\tilde{\ell}^k_A\tilde{\ell}'^k_B}_{\tilde{q}^i_C\tilde{q}^j_D}\left(Q^2\right)\text{Re}\,C_{00}(0,Q^2,0,m_{\tilde{g}}^2,m_{\tilde{q}^i_C}^2,m_{\tilde{q}^j_D}^2;\mu_R)\delta\left(1-z\right)\notag\\
    &\qquad+\frac{1}{2}\sum_{n=i,j}\sum_{C=1}^2F^{\tilde{\ell}^k_A\tilde{\ell}'^k_B}_{\tilde{q}^n_Cq_iq_j}\left(Q^2\right)\text{Re}\,B_1(0,m_{\tilde{g}}^2,m_{\tilde{q}^n_C}^2;\mu_R)\delta\left(1-z\right)\text{,}
    \label{eq:qq_NLO}\\
    \frac{w_{q_ig}^{\text{NLO}}}{T_F\frac{\alpha_s}{2\pi}} &= \sum_jF^{\tilde{\ell}^k_A\tilde{\ell}'^k_B}_{q_iq_j}\left(Q^2\right)\left[\left(\left(1-z\right)^2+z^2\right)\left(\log{\frac{\left(1-z\right)^2}{z}}-\log{\frac{\mu_F^2}{Q^2}}\right)+\frac{1}{2}+3z-\frac{7}{2}z^2\right]\text{,}
    \label{eq:qg_NLO}\\
    w_{\bar q_ig}^{\text{NLO}} &= w_{q_ig}^{\text{NLO}}, \\
    w_{q_i\bar{q}_j}^{\text{match}} &= \frac{1}{2\pi i}\int_{c-i\infty}^{c+i\infty}dNz^{-N}\left(W^{\text{res}}_{q_i\bar{q}_j}-W^{\text{exp}}_{q_i\bar{q}_j}\right)\text{,}
    \label{eq:wqq_match}\\
    w_{q_ig}^{\text{match}} &= \frac{1}{2\pi i}\int_{c-i\infty}^{c+i\infty}dNz^{-N}\left(W^{\text{res}}_{q_ig}-W^{\text{exp}}_{q_ig}\right)\text{,}
    \label{eq:wqg_match}\\
    w_{\bar q_ig}^{\text{match}} &= w_{q_ig}^{\text{match}} \text{.}
\end{align}
Here $C_{00}$ and $B_1$ are Passarino--Veltman coefficients~\cite{Passarino:1978jh}, defined according to the normalization used in \textsc{LoopTools}~\cite{Hahn_1999,vanOldenborgh:1989wn}. These appear as a result of the supersymmetric QCD contribution. For easy reference the $\text{SU}(3)$ group theory factors used are $T_F=\frac{1}{2}$, $C_A=N_C=3$, and $C_F=\frac{N_C^2-1}{2N_C}=\frac{4}{3}$.

The various $F$-functions above are ``effective couplings'' given by:\footnote{In all cases, $F^{\tilde{\ell}^k_A\tilde{\ell}'^k_B} = F^{\tilde{\ell}'^k_B\tilde{\ell}^k_A}$.}
\begin{align}
    F^{\tilde{\ell}^k_A\tilde{\ell}^k_B}_{q_iq_j}\left(Q^2\right) &= \left[Q_{q_i}^2Q_{\tilde{\ell}}^2\delta_{AB}+2Q_{q_i}Q_{\tilde{\ell}}\delta_{AB}\text{Re}\left\{\frac{Z^{AB}_{\tilde{\ell}^k}\left(Z_{q^i_L}+Z_{q^i_R}\right)}{s_W^2c_W^2}\frac{Q^2}{Q^2-\mu_Z^2}\right\}\vphantom{\frac{\left(Z^{AB}_{\tilde{\ell}^k}\right)^2\left(Z_{q^i_L}^2+Z_{q^i_R}^2\right)}{s_W^4c_W^4}}\right.\notag\\
    &\qquad\left.+2\frac{\left\lvert Z^{AB}_{\tilde{\ell}^k}\right\rvert^2\left(\left\lvert Z_{q^i_L}\right\rvert^2+\left\lvert Z_{q^i_R}\right\rvert^2\right)}{\left\lvert s_W^2\right\rvert^2\left\lvert c_W^2\right\rvert^2}\frac{Q^4}{\left\lvert Q^2-\mu_Z^2\right\rvert^2}\right]\delta_{ij}\text{,}\\
    F^{\tilde{l}^k_A\tilde{\nu}^k_B}_{q_iq_j}\left(Q^2\right)  &= \frac{\left\lvert V_{q_iq_j}\right\rvert^2\left(W^{AB}_{\tilde{l}^k\tilde{\nu}^k}\right)^2Q^4}{8\left\lvert s_W^2\right\rvert^2\left\lvert Q^2-\mu_W^2\right\rvert^2}\text{,}\\
    F^{\tilde{\ell}^k_A\tilde{\ell}^k_B}_{\tilde{q}^i_C\tilde{q}^j_D}\left(Q^2\right) &= \left[Q_{\tilde{q}_i}^2Q_{\tilde{\ell}}^2\delta_{AB}\delta_{CD}\vphantom{\frac{\left(Z^{AB}_{\tilde{L}^k}\right)^2}{s_W^4c_W^4}}\right.\notag\\
    +Q_{\tilde{q}_i}Q_{\tilde{\ell}}&\delta_{AB}\delta_{CD}\left.\text{Re}\left\{\frac{Z^{AB}_{\tilde{\ell}^k}}{s_W^2c_W^2}\frac{Q^2}{Q^2-\mu_Z^2}\left(Z^{CD}_{\tilde{q}^i}+2\mathbf{q}^i_{C1}\mathbf{q}^j_{D1}Z_{q_L}+2\mathbf{q}^i_{C2}\mathbf{q}^j_{D2}Z_{q_R}\right)\right\}\right.\notag\\
    +2&\left.\frac{\left\lvert Z^{AB}_{\tilde{\ell}^k}\right\rvert^2}{\left\lvert s_W^2\right\rvert^2\left\lvert c_W^2\right\rvert^2}\frac{Q^4}{\left\lvert Q^2-\mu_Z^2\right\rvert^2}\text{Re}\left\{Z^{CD*}_{\tilde{q}^i}\left(\mathbf{q}^i_{C1}\mathbf{q}^j_{D1}Z_{q_L}+\mathbf{q}^i_{C2}\mathbf{q}^j_{D2}Z_{q_R}\right)\right\}\right]\delta_{ij}\text{,}\\
    F^{\tilde{l}^k_A\tilde{\nu}^k_B}_{\tilde{q}^i_C\tilde{q}^j_D}\left(Q^2\right) &= \frac{V_{q_iq_j}\left(W^{AB}_{\tilde{l}^k\tilde{\nu}^k}\right)^2Q^4}{4\left\lvert s_W^2\right\rvert^2\left\lvert Q^2-\mu_W^2\right\rvert^2}W^{CD*}_{\tilde{q}^i\tilde{q}^j}\mathbf{q}^i_{C1}\mathbf{q}^j_{D1}\text{,}\\
    F^{\tilde{\ell}^k_A\tilde{\ell}^k_B}_{\tilde{q}^n_Cq_iq_j}\left(Q^2\right) &= \left[\frac{1}{2}Q_{q_n}^2Q_{\tilde{\ell}}^2\delta_{AB}^2\vphantom{\frac{\left(Z^{AB}_{\tilde{\ell}^k}\right)^2}{s_W^4c_W^4}}\right.\notag\\
    +2&\left.Q_{q_n}Q_{\tilde{\ell}}\delta_{AB}\text{Re}\left\{\frac{Z^{AB}_{\tilde{\ell}^k}}{s_W^2c_W^2}\frac{Q^2}{Q^2-\mu_Z^2}\left(\left(\mathbf{q}^n_{C1}\right)^2Z_{q_L}+\left(\mathbf{q}^n_{C2}\right)^2Z_{q_R}\right)\right\}\right.\notag\\
    +2&\left.\frac{\left\lvert Z^{AB}_{\tilde{\ell}^k}\right\rvert^2}{\left\lvert s_W^2\right\rvert^2\left\lvert c_W^2\right\rvert^2}\frac{Q^4}{\left\lvert Q^2-\mu_Z^2\right\rvert^2}\left(\left(\mathbf{q}^n_{C1}\right)^2\left\lvert Z_{q_L}\right\rvert^2+\left(\mathbf{q}^n_{C2}\right)^2\left\lvert Z_{q_R}\right\rvert^2\right)\right]\delta_{ij}\\
    F^{\tilde{l}^k_A\tilde{\nu}^k_B}_{\tilde{q}^n_Cq_iq_j}\left(Q^2\right) &= \frac{\left\lvert V_{q_iq_j}\right\rvert^2\left(W^{AB}_{\tilde{l}^k\tilde{\nu}^k}\right)^2Q^4}{8\left\lvert s_W^2\right\rvert^2\left\lvert Q^2-\mu_W^2\right\rvert^2}\left(\mathbf{q}^n_{C1}\right)^2\text{.}
\end{align}
Here $Q_i$ is the electric charge of particle $i$; the ``complex masses'' $\mu_V$ are defined in \eqref{eq:cms}; $s_W,c_W=\sin\theta_W,\cos\theta_W$ with complex values as defined in \eqref{eq:sintheta_cms}; $V$ is the CKM matrix; and the various couplings of $Z$ and $W$ bosons to quarks, sleptons and squarks, are given by
\begin{align}
    Z_{q^i_{L/R}} &= -\frac{1}{2}\left(T^3_{q^i_{L/R}}-Q_{q_i}s_W^2\right)\text{,}\\
    Z^{AB}_{\tilde{\ell}^k} &= -\left(T^3_\ell\boldsymbol{\ell}^k_{A1}\boldsymbol{\ell}^k_{B1}-\delta_{AB}Q_{\tilde{\ell}}s_W^2\right)\text{,}\\
    Z^{CD}_{\tilde{q}^i} &= -\left(T^3_{q^i}\mathbf{q}^i_{C1}\mathbf{q}^i_{D1}-\delta_{CD}Q_{\tilde{q}^i}s_W^2\right)\text{,}\\
    W^{AB}_{\tilde{l}^k\tilde{\nu}^k} &= \mathbf{l}^k_{A1}\boldsymbol{\nu}^k_{B1}\text{,}\\
    W^{AB}_{\tilde{q}^i\tilde{q}^j} &= \mathbf{q}^i_{A1}\mathbf{q}^j_{B1}V_{q_iq_j}\text{,}
\end{align}
where $T_i^3$ is the third component of weak isospin for particle $i$, and where $\mathbf{f}^k_{AB}$ is the mixing matrix for sfermions given in Eq.~(\ref{eq:slepton_mixing_matrix}). 

Lastly, the resummation coefficients of \eqref{eq:wqq_match} and \eqref{eq:wqg_match} are given by
\begin{equation}
    W^{\text{res}}_{q_i\bar{q}_j} = F^{\tilde{\ell}^k_A\tilde{\ell}'^k_B}_{q_iq_j}\left(Q^2\right)\exp{\left\{\log{\bar{N}}g_1\left(\lambda\right)+g_2\left(\lambda\right)+h_1\left(\lambda,N\right)\right\}}\text{,}
    \label{eq:qqres_app}
\end{equation}
with~\cite{Catani:2003zt,vanBeekveld:2021hhv}
\begin{align}
    g_1\left(\lambda\right) &= \frac{A_1}{\pi b_0\lambda}\left(2\lambda+\left(1-2\lambda\right)\log{\left(1-2\lambda\right)}\right)\text{,}\label{eq:g1}\\
    g_2\left(\lambda\right) &= -\frac{A_2}{\pi^2b_0^2}\left(2\lambda+\log{\left(1-2\lambda\right)}\right)+\frac{A_1b_1}{\pi b_0^3}\left(2\lambda+\log{\left(1-2\lambda\right)}+\frac{1}{2}\log^2{\left(1-2\lambda\right)}\right)\notag\\
    &\qquad-\frac{A_1}{\pi b_0}\left(2\lambda+\log{\left(1-2\lambda\right)}\right)\log{\frac{\mu_R^2}{Q^2}}+\frac{2A_1\lambda}{\pi b_0}\log{\frac{\mu_F^2}{Q^2}}\text{,}
    \label{eq:g2}\\
    h_1\left(\lambda,N\right) &= -\frac{A_1}{\pi b_0}\frac{\log{\left(1-2\lambda\right)}}{N}\text{.}
    \label{eq:h1}
\end{align}
Here $\bar{N}\equiv Ne^{\gamma_E}$ and $\lambda \equiv b_0\alpha_s\log{\bar{N}}$, with constants~\cite{Vogt:2004mw,Moch:2004pa,Gross:1973id,Politzer:1973fx,Caswell:1974gg,Jones:1974mm,Egorian:1978zx}
\begin{align}
    4\pi b_0 &= \frac{11}{3}C_A-\frac{4}{3}T_FN_F\text{,}\\
    \left(4\pi\right)^2b_1 &= \frac{34}{3}C_A^2-\frac{20}{3}C_AT_FN_F-4C_FT_FN_F\text{,}\\
    A_1 &= C_F\text{,}\\
    A_2 &= C_F\left(C_A\left(\frac{67}{36}-\frac{\pi^2}{12}\right)-N_F\frac{5}{18}\right)\text{,}
\end{align}
with $N_F=5$ approximately massless quark flavors.

Next, we have~\cite{vanBeekveld:2021mxn}
\begin{align}
    W^{\text{res}}_{q_ig}& = \sum_jF^{\tilde{\ell}^k_A\tilde{\ell}'^k_B}_{q_iq_j}\left(Q^2\right)\frac{T_F}{C_A-C_F}\frac{1}{2N\log{N}}\notag\\
    &\times\left[\exp{\left\{2\frac{\alpha_s}{\pi}C_F\log^2{N}\right\}}\mathcal{B}_0\left(\frac{\alpha_s}{\pi}\left(C_A-C_F\right)\log^2{N}\right)-\exp{\left\{\frac{\alpha_s}{2\pi}\left(C_F+3C_A\right)\log^2{N}\right\}}\right]\text{,}
\end{align}
where 
\begin{equation}
    \mathcal{B}_0\left(x\right) \equiv 1-\frac{x}{2}-\sum_{n=1}^\infty\frac{\left(-1\right)^n}{\left(2n!\right)^2}\left\lvert B_{2n}\right\rvert x^{2n}\text{,}
\end{equation}
with $B_n$ the Bernoulli numbers.

The ``expanded'' terms of \eqref{eq:wqq_match} and \eqref{eq:wqg_match} that subtract the fixed order contributions are
\begin{align}
    W^{\text{exp}}_{q_i\bar{q}_j} &= F^{\tilde{\ell}^k_A\tilde{\ell}'^k_B}_{q_iq_j}\left(Q^2\right)\left[1+C_F\frac{\alpha_s}{\pi}\left(2\log^2{\bar{N}}+2\log{\frac{\mu_F^2}{Q^2}}\log{\bar{N}}+2\frac{\log{\bar{N}}}{N}\right)\right]\text{,}\\
    W^{\text{exp}}_{q_ig} &= -\sum_jF^{\tilde{\ell}^k_A\tilde{\ell}'^k_B}_{q_iq_j}\left(Q^2\right)T_F\frac{\alpha_s}{\pi}\frac{\log{N}}{N}\text{.}
\end{align}

\section{Convergence of the inverse Mellin transform}
\label{sec:Mellin transform}
Here we will discuss the convergence of the inverse Mellin transform that is used in this paper in some detail. We begin by introducing some basic analytical properties of the Mellin transform before specializing to the specific problem faced in the calculation of (resummed) cross sections. Throughout, since fixed-order partonic cross sections are distributions, we adopt a distributional framework and interpret Mellin transforms as pairings with test functions.

\subsection{Analytical properties of the Mellin transform}
\label{eq:analytical_properties_Mellin}
As the Mellin transform is closely related to the Laplace transform, it is instructive to first examine the conditions under which the Laplace transform converges. Let us begin with the notion of local integrability. A function $f(x)\in L_{\textnormal{loc}}^{1}(\Omega)$, $\Omega$ possibly an open set, is said to be locally integrable iff for every compact set $K\subset \Omega$, the Lebesgue integral
\begin{align}
    \int_{K}dx\,|f(x)|\,<\,\infty\,, 
\end{align}
is finite. This condition ensures that the function behaves well on every bounded portion of its domain, even if the overall domain is unbounded.

In many settings---especially when dealing with integral transforms---it is not enough to know that a function is locally integrable; one must also control the behavior at infinity or near the boundary of the domain. This is where the idea of growth condition comes into play. By a growth condition, we mean a restriction on how rapidly the function grows outside any compact set. When a function $f(x)$ is locally integrable and its growth is appropriately controlled by these conditions, we can extend local integrability to absolute integrability. In other words, under the right growth restrictions, the function satisfies
\begin{align}
    \int_{\Omega}dx\,|f(x)|<\infty\,,
\end{align}
which means that the function $f$ is globally Lebesgue integrable, i.e.\ $f\in L^{1}(\Omega)$.

For instance, if a function $g(t)\in L_{\textnormal{loc}}^{1}([0,\infty))$ satisfies the bound\footnote{On finite intervals $[0,a]$ local integrability coincides with absolute integrability on that interval. We could also consider the interval $t\in(0,\infty)$, but then we would need an additional condition when $t\rightarrow 0$.}
\begin{align}
    \int_{0}^{a}dt\,|g(t)|\,<\,\infty\,, 
\end{align}
for any finite $a>0$ {\em and} if $g(t)=\mathcal{O}(e^{\alpha t})$ 
as $t\rightarrow\infty$ for some real constant $\alpha$,\footnote{Meaning $g(t)$ grows no faster than a constant times $e^{\alpha t}$.} it has a one-sided Laplace transform
\begin{align}
    \mathcal{L}^{+}\{g\}(N)=\int_{0}^{\infty}dt\,g(t)e^{-Nt}:=G(N)\,,
\end{align}
which converges absolutely and is analytic in the right-half plane $\textnormal{Re}(N)=\sigma>\alpha$. 

That is, local integrability only ensures that $g(t)$ does not misbehave locally---say, near $t=0$. Without a bound on the growth as $t\rightarrow\infty$, the tail of the function might be too large for the integral to converge when taken over $[0,\infty)$. Thus, we suppose there exists a constant $M>0$ and $\alpha\in\mathbb{R}$ such that
\begin{align}
    |g(t)|\leq Me^{\alpha t}\,,\hspace{0.2cm}\forall t\geq 0\,,
\end{align}
giving that for $\sigma >\alpha$
\begin{align}
    |G(N)|=\Big|\int_{0}^{\infty}dt\, g(t)e^{-Nt}\Big|&\leq \int_{0}^{\infty}dt\,|g(t)|e^{-\sigma t}
    \leq M\int_{0}^{\infty}dt\,e^{-(\sigma-\alpha)t}
    =\frac{M}{\sigma - \alpha} <\infty\,,
\end{align}
and so we see that the Laplace transform converges absolutely in the half-plane $\sigma >\alpha$, giving that the region of convergence is $(\alpha,\infty)$. 



Since the Laplace transform converges absolutely in the half-plane $ \sigma > \alpha $, it follows by standard arguments (Weierstrass M-test, dominated convergence theorem) that the convergence is uniform on every compact subset of this region. Moreover, for each fixed $t$, the function $g(t)e^{-Nt}$ is entire in $N$ (its power series expansion converges for all $N$). Since the integrand is entire in $N$ and the convergence is uniform on compact sets, we are justified in differentiating under the integral sign, i.e.\ the Laplace transform can be differentiated term by term with respect to $N$. Consequently, it is infinitely differentiable with respect to $N$, and by classical results in complex analysis (such as Morera's theorem), the Laplace transform is analytic in the region $ \sigma > \alpha $.


Similarly, suppose $g(t)\in L_{\textnormal{loc}}^{1}((-\infty,0])$, giving that
\begin{align}
    \int_{0}^{a}dt\,|g(-t)|\,<\,\infty\,, 
\end{align}
and $g(t)=\mathcal{O}(e^{\beta t})$ as $t\rightarrow -\infty$ for some real constant $\beta$, the one-sided Laplace transform
\begin{align}
    \mathcal{L}^{-}\{g\}(N)=\int_{-\infty}^{0}dt\,g(t)e^{-Nt}=\int_{0}^{\infty}dt\,g(-t)e^{Nt},
\end{align}
converges absolutely and is analytic in the left half plane $\textnormal{Re}(N)<\beta$, i.e. region of convergence is $(-\infty,\beta)$. 

Further, consider $g(t)\in L_{\textnormal{loc}}^{1}(\mathbb{R})$ with the growth conditions $g(t)=\mathcal{O}(e^{\alpha t})$ as $t\rightarrow \infty$ and $g(t)=\mathcal{O}(e^{\beta t})$ as $t\rightarrow -\infty$, and if $\alpha <\beta$ the two-sided Laplace transform 
\begin{align}
    \mathcal{L}\{g\}(N)=\int_{-\infty}^{\infty}dt\,g(t)e^{-Nt}
\end{align}
converges absolutely and is analytic in the vertical strip $\alpha<\textnormal{Re}(N)<\beta$. 

\smallskip
There is an intrinsic connection between the Laplace and Mellin transforms that can be unveiled through an appropriate change of variables: If we consider $x=e^{-t}$, where $t\in [0,\infty)$, so that $t=-\ln x$ and $dt=-dx/x$, the one-sided Laplace transform can be written as
\begin{align}
    \mathcal{L}_{[0,\infty)}\{g\}(N)=\int_{0}^{\infty}dt\, g(t)e^{-Nt}=\int_{0}^{1}dx\,x^{N-1}g(-\ln x),
\end{align}
which is precisely the restricted Mellin transform of the function $f(x):=g(-\ln x)$ for $x\in(0,1]$
\begin{align}\label{eq:mellinoneside}
    \mathcal{M}_{(0,1]}\{f\}(N)=\int_{0}^{1}dx\,x^{N-1}f(x)\,.
\end{align}
The conditions on $g(t)$ translate to $f(x)=\mathcal{O}(x^{-\alpha})$ as $x\rightarrow 0^+$ and that $\int_{b}^{1}dx|f(x)|<\infty$ for some finite $b<1$, such that the integral converges absolutely and is analytic in the right-half plane $\textnormal{Re}(N)>\alpha$.
Similarly, for the two-sided Laplace transform we get a standard Mellin transform for $x\in(0,\infty)$
\begin{align}\label{eq:mellintwoside}
    \mathcal{L}_{(-\infty,\infty)}\{g\}(N)&=\int_{-\infty}^{\infty}dt\,g(t)e^{-Nt}
    =\int_{0}^{\infty}dx\,x^{N-1}f(x)
    =\mathcal{M}_{(0,\infty)}\{f\}(N)\,,
\end{align}
and the growth conditions on $g(t)$ translate to the Mellin transform as:
\begin{enumerate}
    \item Since $g(t)=\mathcal{O}(e^{\alpha t})$ as $t\rightarrow\infty$ we obtain $f(x)=\mathcal{O}(x^{-\alpha})$ as $x\rightarrow 0^{+}$.
    \item Since $g(t)=\mathcal{O}(e^{\beta t})$ as $t\rightarrow -\infty$ we obtain $f(x)=\mathcal{O}(x^{-\beta})$ as $x\rightarrow \infty$.
\end{enumerate}
which ensure that the integral converges absolutely and is analytic in the vertical strip $\alpha <\textnormal{Re}(N)<\beta$. 

A standard example is the function $f(x)=e^{-x}$, satisfying $e^{-x}=\mathcal{O}(x^{0})$ as $x\rightarrow 0^{+}$ and $e^{-x}=\mathcal{O}(x^{-b})$ as $x\rightarrow\infty$ for any $b>0$, so that its Mellin transform
\begin{align}
    F(N)=\int_{0}^{\infty}dx\, x^{N-1}e^{-x}=\Gamma(N)\,,
\end{align}
the Gamma function, is defined and analytic in the right half-plane $\textnormal{Re}(N)>0$.

\bigskip
To obtain the inversion formula for the Mellin transform, we use that the two-sided Laplace transform has the inverse transform
\begin{align}
    g(t)=\frac{1}{2\pi i}\int_{c-i\infty}^{c+i\infty}dN\,G(N)e^{Nt}.
\end{align}
The integration is along a vertical line through $\textnormal{Re}(N)=c$ in the complex $N$ plane, known as the Bromwich contour. Then, again setting $t=-\ln x$ and $f(x)=g(-\ln x)$, we obtain the inverse Mellin transform
\begin{align}\label{eq:MellinInv}
    f(x)=\frac{1}{2\pi i}\int_{c-i\infty}^{c+i\infty}dN\,x^{-N}F(N).
\end{align}

If the Mellin transform $F(N)$ is analytic in the strip $(\alpha,\beta)$ and vanishes sufficiently fast when $\textnormal{Im}(N)\rightarrow\pm\infty$, then by Cauchy's theorem, the path of integration can be translated sideways and deformed inside the strip without affecting the result of the integration. More precisely, for $F(N)$ analytic in the strip $(\alpha,\beta)$ and which satisfies
\begin{align}\label{eq:inequality}
    |F(N)|\leq K|N|^{-2}\,,
\end{align}
for some constant $K>0$, the function $f(x)$ obtained by \eqref{eq:MellinInv} is a continuous function and its Mellin transform is $F(N)$. Although this gives a sufficient condition for the inversion formula to yield a continuous function, in practice it is important to emphasise that the inversion formula applies to a function $F$ analytic in a given strip and that the uniqueness of the result holds only with respect to that strip. In other words, a Mellin transform consists of a pair: a function $F(N)$ and a strip of analyticity $(\alpha,\beta)$. Thus, a unique function $F(N)$ with several disjoint strips of analyticity will generally have several Mellin inverses, one for each strip; a Mellin transform is well defined only when its strip of analyticity is specified.

To get a feel for the inverse, let us look at the inverse of the gamma function $F(N)=\Gamma(N)$ considered above and show that we obtain $f(x)=e^{-x}$. From Stirling's formula we have the growth condition
\begin{align}
    |\Gamma(c+iy)|\sim \sqrt{2\pi}|y|^{c-1/2}e^{-\frac{\pi}{2}|y|}\,,\hspace{0.2cm}\text{as}\,\,\,|y|\rightarrow\infty\,,
\end{align}
such that \eqref{eq:inequality} is satisfied and the inversion formula \eqref{eq:MellinInv} can be applied. The gamma function is analytic in the half-plane $\textnormal{Re(N)}>0$, with an infinite number of isolated poles at the negative or zero integers. Thus, it admits an analytic continuation to a meromorphic function on $\mathbb{C}$ with simple poles at $N=-n\in -\mathbb{N}_0$ of residue
\begin{align}
    \textnormal{Res}[\Gamma(N),N=-n]=\frac{(-1)^n}{n!}\,.
\end{align}
By choosing a rectangular contour, where the contributions from the horizontal parts vanish as $\textnormal{Im}(N)$ goes to infinity, the integral will pick up the values of the residues at each enclosed pole. Specifically, if $c>0$ and $-k< c' <-k+1$, $k\in\mathbb N$, we have
\begin{align}
    \frac{1}{2\pi i}\int_{c-i\infty}^{c+i\infty}dN\,x^{-N}\Gamma(N) &=\sum_{n=0}^{k-1}\textnormal{Res}[x^{-N}\Gamma(N),N=-n]+\frac{1}{2\pi i}\int_{c'-i\infty}^{c'+i\infty}dN\,x^{-N}\Gamma(N)\nonumber
    \\
    &=\sum_{n=0}^{k-1}\frac{(-1)^n}{n!}x^{n}+\frac{1}{2\pi i}\int_{c'-i\infty}^{c'+i\infty}dN\,x^{-N}\Gamma(N).
\end{align}
Here, the sum is the $k$ first terms in the Taylor series for $e^{-x}$ and the integral term represents the remainder, which can be shown to vanish in the limit $k\rightarrow\infty$ by the Stirling formula, such that
\begin{align}
    \frac{1}{2\pi i}\int_{c-i\infty}^{c+i\infty}dN\, x^{-N}\Gamma(N)=e^{-x}\,.
\end{align}
Thus, for functions admitting an analytic continuation to a meromorphic function with (possibly an infinite number of) simple poles, we can easily apply standard methods from complex analysis to calculate the inverse Mellin transform. However, for functions possessing branch points, special treatment is needed. Recall, a branch point is a location in the complex plane where the function inherently becomes multi-valued. In any neighborhood of such a point, it is impossible to define the function in a single, continuous manner because if one encircles the branch point, the function will generally assume a different value upon returning to the starting point. To handle this, one typically introduces a branch cut --- a carefully chosen curve or line extending from the branch point (or between branch points) to infinity --- that effectively \textquote{cuts} the plane. This cut restricts the domain so that the function becomes single-valued and analytic on the resulting region.


The consequence for the present discussion is that the inverse Mellin is not well-defined without generalizing the formulation. That is, branch points in Mellin space reflect non-analytic endpoint behaviour and, if sufficiently strong, may lead to a non-integrable structure in $x$-space. In addition, there are certainly well-known cases where the growth condition \eqref{eq:inequality} on the Mellin transform is not satisfied.  It is therefore necessary to generalize the Mellin transform to cases where the function $f(x)$ is too singular to admit the usual Mellin transforms we defined above, and to handle possibly severe growth condition of $F(N)$ preventing a naive inverse transform. A rigorous way of handling this is to resort to distribution theory.

\bigskip
In distribution theory, \textquote{generalized functions} extend the notion of classical functions to include objects that are not locally integrable, which cannot be defined pointwise but can be interpreted as continuous linear functionals acting on a space of test functions. They are used to give meaning to otherwise ill-defined operations. To clarify the usefulness of distribution theory, we consider the following classification
\begin{enumerate}
    \item If $f(x)$ is locally integrable on an open set $D\subset \mathbb R$, we define a \emph{regular} distribution through the integral\footnote{This is not an inner product, just a way of denoting how the distribution acts on the test function.}
    \begin{align}
        \langle T_f,\phi\rangle=\int_{D} dx\, f(x)\phi(x),
    \end{align}
    where $\phi$ is an appropriate test function, with compact support or of rapid decay at infinity. Notationally, these spaces are often denoted as $\mathcal{D}(\mathbb{R})$ and $\mathcal{S}(\mathbb{R})$, respectively.
    \item If $f(x)$ is not locally integrable on $D$, we can define a \emph{singular} distribution through a limiting procedure: introduce a family $\{f_\epsilon\}$ of functions locally integrable on $D_{\epsilon}\subset D$, a domain with the singular region (or regions) excised, and define the distributional limit
    \begin{align}
        \langle T_f,\phi\rangle=\lim_{\epsilon\rightarrow 0}\Big(\int_{D_{\epsilon}}dx\, f_{\epsilon}(x)\phi(x)+C_{\epsilon}[\phi]\Big),
    \end{align}
    where $C_{\epsilon}[\phi]$ is an appropriate counterterm of distributions supported at the singular point, denoting the possible ambiguity in the construction.
    
\end{enumerate}
Let us explain the construction of singular distributions with a couple of examples: Consider the function $f(x)=1/x$. Clearly, because of the singularity at $x=0$, this is not a locally integrable function on $\mathbb{R}$. We want to know if it is possible to construct a distribution out of $1/x$. Let us formalize this in the following way: $f(x) = 1/x$ is obviously a locally integrable function on $\mathbb{R}\setminus\{0\}$. Thus, for $
\phi\in\mathcal{D}(\mathbb{R}\setminus\{0\})$,  $T_{1/x}$ is a distribution $\mathcal{D}'({\mathbb{R}\setminus \{0\}})$. Now, we want to find a distribution on $\mathbb{R}$ that coincides with $1/x$ on $\mathbb{R}\setminus\{0\}$. The following are instances of such a construction:
    \begin{enumerate}
        \item The principal value
        prescription: we can construct a distribution out of $1/x$ by using principal value integrals \begin{align}
         \langle\textnormal{p.v.}\Big[\frac{1}{x}\Big],\phi\rangle&=\textnormal{p.v.}\int_{-\infty}^{\infty}dx\,\frac{\phi(x)}{x}\,,
    \end{align}
    where
    \begin{align}
        \textnormal{p.v.}\int_{-\infty}^{\infty}dx\,\frac{\phi(x)}{x}&=\lim_{\varepsilon\to0^{+}}
        \left(
            \int_{-\infty}^{-\varepsilon}\frac{\phi(x)}{x}\,dx
            +\int_{\varepsilon}^{\infty}\frac{\phi(x)}{x}\,dx
        \right)\nonumber
        \\
        &=\int_{0}^{\infty}\frac{\phi(x)-\phi(-x)}{x}\,dx\,,
    \end{align}
    corresponds to $1/x$ on $\mathbb{R}\setminus\{0\}$ and defines a distribution $\mathcal{D}'(\mathbb{R})$ for $
    \phi\in\mathcal{D}(\mathbb{R})$. Note here that no counterterm is needed as the principal value uses symmetric integration around the singular point.
    \item The Feynman prescription\footnote{Not the standard name in mathematics, but used for similar analytic continuation in physics.}: If $\epsilon >0$ then the functions $f_{\epsilon}=1/(x\pm i\epsilon)$ is locally integrable and therefore defines a regular distributions in $\mathcal{D}(\mathbb{R})$. If we now let $\epsilon\rightarrow 0$ we obtain the singular distributions
    \begin{align}
        \langle\Big[ \frac{1}{x\pm i0}\Big],\phi\rangle&=\lim_{\epsilon\rightarrow 0}\int_{-\infty}^{\infty}dx\,\frac{\phi(x)}{x\pm i\epsilon}\,,
    \end{align}
    which also corresponds to $1/x$ on $\mathbb{R}\setminus\{0\}$ and defines a distribution $\mathcal{D}'(\mathbb{R})$ for $
    \phi\in\mathcal{D}(\mathbb{R})$.
    \item Another useful strategy in defining a singular distribution is the following: define a function $f(x)$ that is locally integrable on $\mathbb{R}\setminus\{0\}$. Then, find a locally integrable function $u(x)$, whose derivative corresponds to $f(x)$ on $\mathbb{R}\setminus\{0\}$. For example, let $f(x)=1/x$ and consider the locally integrable function $u(x)=\ln |x|$. The derivative $u'(x)$ corresponds to $1/x$ on $\mathbb{R}\setminus\{0\}$ and is a singular distribution since
    \begin{align}
        \frac{d}{dx}\ln|x|=\textnormal{p.v.}\Big(\frac{1}{x}\Big)\,,
    \end{align}
    which is found by calculating the distributional derivative
    \begin{align}
        \langle \frac{d}{dx}\ln|x|,\phi\rangle&=-\langle \ln|x|,\phi'\rangle\nonumber
        \\
        &=-\int_{-\infty}^{\infty}dx\,\ln |x|\frac{d}{dx}\phi(x)\nonumber
        \\&=\lim_{\varepsilon\to0^{+}}
   \Bigl(
       \int_{-\infty}^{-\varepsilon}\frac{\phi(x)}{x}\,dx
       +\int_{\varepsilon}^{\infty}\frac{\phi(x)}{x}\,dx
       \;-\;
       \ln\varepsilon\,\bigl[\phi(-\varepsilon)-\phi(\varepsilon)\bigr]
   \Bigr)\nonumber
   \\
       &=\langle \textnormal{p.v.}\Big(\frac{1}{x}\Big),\phi\rangle\,.
    \end{align}
    where $\ln\epsilon(\phi(-\epsilon)-\phi(\epsilon))$ vanishes in the $\epsilon\rightarrow0$ limit.
\end{enumerate} 
It is quite clear that none of these are functions, and the reason is that all are equal to $1/x$ on $\mathbb{R}\setminus\{{0}\}$. In other words, singular distributions are not uniquely defined, and a simple calculation shows that\footnote{This is the well-known Sokhotski-Plemelj formula.}
\begin{align}
    \frac{1}{x\pm i0} &= \operatorname{p.v.}\left(\frac{1}{x}\right) \mp i\pi\,\delta(x)\,.
\end{align}
This is a particular instance of the general fact that if two distributions \(T_1\) and \(T_2\) agree on \(\mathbb{R}\setminus\{0\}\), then their difference is a distribution whose support is contained in \(\{0\}\). By Schwartz’s structure theorem, any distribution supported at a single point must be a finite linear combination of the Dirac delta function and possibly---if the singular structure is severe---its derivatives. The construction considered here goes under the name of \emph{extension of distributions}. In other words, one can in cases (when the singular structure is not too severe) extend a distribution defined on $\mathcal{D}'(\mathbb{R}\setminus\{0\})$ to a distribution on $\mathcal{D}'(\mathbb{R})$, though this extension is not necessarily unique.

Another class of useful extensions are those where the singularities are at the boundaries, where symmetric extensions cannot be used. A standard example is the \emph{Hadamard partie finie} distribution
\begin{align}
    \langle \textnormal{Pf}\Big(\frac{\theta(x)}{x}\Big),\phi\rangle=\textnormal{F.p.}\int_{0}^{\infty}dx\frac{\phi(x)}{x},
\end{align}
also called the finite part integral, where one uses that
\begin{align}
    \int_{\epsilon}^{\infty}dx\,\frac{\phi(x)}{x}=-\phi(0)\ln\epsilon+\int_{\epsilon}^{1}dx\frac{\phi(x)-\phi(0)}{x}+\int_{1}^{\infty}dx\frac{\phi(x)}{x}\,,
\end{align}
such that the limit $\epsilon\rightarrow 0$ can be taken, giving
\begin{align}
    \textnormal{F.p.}\int_{0}^{\infty}dx\frac{\phi(x)}{x}=\int_{0}^{1}dx\frac{\phi(x)-\phi(0)}{x}+\int_{1}^{\infty}dx\frac{\phi(x)}{x}\,.
\end{align}
In this case, one can also find the distributional identity
\begin{align}
    \frac{d}{dx}(\theta(x)\ln x)=\textnormal{Pf}\Big(\frac{\theta(x)}{x}\Big)\,.
\end{align}
Following the singular distribution classification above, this may also be viewed as defining the \emph{extension} $x_{+}^{-1}$ of $\theta(x)x^{-1}$ as
\begin{align}
    x_{+}^{-1}=\lim_{\epsilon\rightarrow 0}\int_{\epsilon}^{\infty}dx\,x^{-1}+\ln\epsilon\, \delta\,,
\end{align}
where we subtracted the distribution $\ln\epsilon\, \delta$ supported at $0$, which becomes
singular when $\epsilon \rightarrow 0$, also called a local counterterm.

With this demonstration in place, let us analyse functions of relevance in this work, i.e.\ functions with endpoint singularities, and study their restricted Mellin transform. We want to make sense of the following functions
\begin{align}
    f(x)&=\frac{1}{1-x},
    \\
    g(x)&=\frac{1}{\ln x}\,,
\end{align}
which are not locally integrable on the domain $(0,1]$, due to the singularity at $x=1$. Therefore, $(1-x)^{-1}$ and $(\ln x)^{-1}$ does not have any restricted (or standard) Mellin transform in the usual sense. For functions with singularities in the interior of the domain, the principal value extension was appropriate, and for endpoints we use Hadamard finite part extensions. Hence, in a singular distributional sense, we define the plus distribution
\begin{align}
    \Big[\frac{1}{1-x}\Big]_{+}:=\textnormal{Pf}\Big(\frac{\theta(1-x)}{1-x}\Big)\,,
\end{align}
as the Hadamard finite part integral
\begin{align}
    \langle\Big[\frac{1}{1-x}\Big]_{+},\phi\rangle=\textnormal{F.p.}\int_{0}^{1}dx\frac{\phi(x)}{1-x}\,,
\end{align}
where one calculates
\begin{align}
    \int_{0}^{1-\epsilon}dx\,\frac{\phi(x)}{1-x}=-\phi(1)\ln\epsilon+\int_{0}^{1-\epsilon}dx\,\frac{\phi(x)-\phi(1)}{1-x}\,,
\end{align}
such that the limit $\epsilon\rightarrow 0$ exist, and we get
\begin{align}
    \langle\Big[\frac{1}{1-x}\Big]_{+},\phi\rangle&
    =\int_{0}^{1}dx\,\frac{\phi(x)-\phi(1)}{1-x}.
\end{align}
Similarly, for later reference, we can also define the singular distribution
\begin{align}
    \Bigl\langle \Bigl[\frac{1}{\ln x}\Bigr]_{+}, \,\phi \Bigr\rangle
    &= \int_{0}^{1}\frac{\phi(x)-\phi(1)}{\ln x}\,dx \,.
\end{align}

The space of distributions forms a natural domain for the Mellin transform. The most practical way to see this is to give a new interpretation to the defining formulas \eqref{eq:mellinoneside} and \eqref{eq:mellintwoside} by
considering them as the action of a distribution $f$\footnote{From now on the distribution is denoted by the function $f$ itself instead of the more precise notation $T_f$.} on a test function $x^{N-1}$ 
\begin{align}\label{eq:mellindist}
    F(N)=\langle f,x^{N-1}\rangle\,,
\end{align}
where a suitable space of test functions $\mathcal{K}(\alpha,\beta)$, containing $x^{N-1}$ for $N$ in the region $\alpha<\textnormal{Re}(N)<\beta$, is the following: The space $\mathcal{K}(\alpha,\beta)$ is composed of functions $\phi(x)$ defined on $(0,\infty)$ and with continuous derivatives of all orders going to zero as $x$ approaches either zero or infinity. More precisely, there exists two positive numbers $\eta_1$, $\eta_2$ such that for all integers $k$, the following conditions hold
\begin{align}
    x^{k+1-\alpha-\eta_1}&\Big(\frac{d}{dx}\Big)^{k}\phi(x)\longrightarrow 0 \,,\hspace{0.2cm}x\rightarrow 0\,,
    \\
    x^{k+1-\beta-\eta_2}&\Big(\frac{d}{dx}\Big)^{k}\phi(x)\longrightarrow 0 \,,\hspace{0.2cm}x\rightarrow \infty \,\,,
\end{align}
with an analogous construction for the restricted Mellin transform.
The space of distributions $\mathcal{K'}(\alpha,\beta)$ is then introduced as a linear space of continuous linear functionals on $\mathcal{K}(\alpha,\beta)$. For locally integrable functions we have regular distributions and the pairing \eqref{eq:mellindist} reduces to the usual Mellin transform. For non-locally integrable functions we may use singular distribution construction. 

With this in place, let us perform the Mellin transform for the singular distributions defined above:
\begin{itemize}
    \item In a plus distributional sense we have the restricted Mellin transform 
    \begin{align}
        \langle \Big[\frac{1}{1-x}\Big]_+,x^{N-1}\rangle=-\psi(N)-\gamma_E\,,\hspace{0.2cm}\textnormal{Re}(N)>0\,,
    \end{align}
    where $\psi(N)$ is the digamma function and $\gamma_E$ is the Euler-Mascheroni constant.
    \item In a plus distributional sense we also have the restricted Mellin transform
    \begin{align}
        \langle \Big[\frac{1}{\ln x}\Big]_+,x^{N-1}\rangle=\ln N \,,\hspace{0.2cm}\textnormal{Re}(N)>0\,,
    \end{align}
    which is found by a change of variable to a standard Laplace integral
    \begin{align}
        \int_{0}^{\infty}dt \frac{e^{-t}-e^{-Nt}}{t}=\ln N\,.
    \end{align}
\end{itemize}
For the study of the inverse Mellin transform of the resummed cross section, the last two are relevant. In $x$-space, the fixed order cross sections are plus distributions. For instance, the dominant (leading-logarithmic) part of the NLO cross section is in $x$-space of the form (up to a numerical factor)
\begin{align}
    \omega_{q\bar{q}}^{\textnormal{LL}}(x)\Big|_{\mathcal{O}(\alpha_s)}=\alpha_s\Bigg(\frac{\ln(1-x)}{1-x}\Bigg)_+\,.
\end{align}
As we are not performing resummation in $x$-space, we need to understand how this expression can be reproduced from a distributional inverse Mellin transform. As we will see in the next section, an exact inverse Mellin transform does not exist for the resummed cross section, but rather as an asymptotic series on the form
\begin{align}
    \sum_{k=0}^{\infty}a_k(\ln N)\alpha_{s}^{k}\,.
\end{align}
where $a_k(\ln N)$ are polynomials in $\ln N$.
For concreteness, let us study the inverse of
$\ln^{k}1/N$.
For $k=1$, we already know that we should reproduce the $[1/\ln x]_+$ distribution. However, we observe that $\ln N$ does not satisfy the growth condition in \eqref{eq:inequality}: the Bromwich contour $N=c+iy$, $y\in(-\infty,\infty)$, $|N|\sim|y|\rightarrow\infty$, and
\begin{align}
    |\ln^{k}N|\rightarrow (\ln|y|)^k\,,
\end{align}
which grows,
and the inverse must in such cases be modified. To restore convergence we use the following result from distribution theory: If there exists an integer $m\geq 0$ such that $N^{2-m}F(N)$ is bounded as $|N|$ goes to infinity, i.e.\ we have the improved bound
\begin{align}
    |N^{-m}F(N)|\leq K|N|^{-2}\,,
\end{align}
then the inverse Mellin transform of $F(N)$ is the unique distribution $f(x)$ given by
\begin{align}
    f(x)=E_{x}^{m}g(x)\,,
\end{align}
where $E_{x}^{m}:=\big(x\partial_x\big)^{m}$ is the Euler operator, and $g(x)$ is a continuous function obtained by the formula
\begin{align}
    g(x)=\frac{(-1)^{m}}{2\pi i}\int_{c-i\infty}^{c+i\infty}dN\,x^{-N}N^{-m}F(N),
\end{align}
where the distributional identity
\begin{align}
    \langle E_{x}^{m}f,x^{N-1}\rangle=(-1)^{m}N^{m}F(N)\,.
\end{align}
has been used, and (as always) $c$ is chosen in the strip of definition. 

Let us put this to work for $F(N)=\ln^{k}1/N$: Along $N=c+iy$ we have $|x^{-N}|=x^{-c}$ and $|N|\sim |y|$ for $y\rightarrow\infty$, so for $m=2$
\begin{align}
    \big|x^{-N} N^{-2}\ln^{k} N\big| \;\lesssim\; x^{-c}\,\frac{(\ln|y|)^{k}}{|y|^{2}}\,,
\end{align}
which is integrable in $y$. Thus, $m=2$ suffices to ensure that the Bromwich integral defining $g$ exists; and, importantly, we may introduce an analytic family 
\begin{align}
    S_{a}=\frac{(-\ln x)^{a-1}}{\Gamma(a)}\,,\hspace{0.4cm}\textnormal{Re(a)}>0\,,
\end{align}
which satisfies $\mathcal{M}\{S_a\}(N)=N^{-a}$, such that
\begin{align}
    g(x)=\frac{(-1)^{m}}{2\pi i}\int_{c-i\infty}^{c+i\infty}dN\, x^{-N} N^{-m}\,
    \partial_a^{k}\!\big(N^{-a}\big)\,\Big|_{a=0}
    \;=\;(-1)^{m}\,\partial_a^{k} S_{a+m}(x)\Big|_{a=0}\,,
\end{align}
where we have used
\begin{align}
    \ln ^{k}\!\frac{1}{N}=\partial_a^{k} N^{-a}\Big|_{a=0}\,,
\end{align}
and where differentiation under the integral sign is justified by dominated convergence for holomorphic families, since the integrand is holomorphic in $a$ and decays sufficiently fast along the Bromwich line
for $m\ge2$, we obtain the exact inversion
\begin{align}
    f(x)=E_x^{m} g(x)=\,\partial_a^{k} S_{a}(x)\Big|_{a=0}\,.
\end{align}

To express this in the standard $x$-space basis, note that for $\textnormal{Re}(a)>0$
\begin{align}
\int_0^1 x^{N-1}\Big[\ln^{a-1}\!\frac{1}{x}\Big]_+\,dx=\Gamma(a)\,(N^{-a}-1)\,,
\end{align}
we obtain the distributional identity
\begin{align}
    S_a(x)=\Delta(a)\Big[\ln^{a-1}\!\frac{1}{x}\Big]_+ + \delta(1-x)\,,
\end{align}
with $\Delta(a)=1/\Gamma(a)$. Differentiating $k\ge1$ times at $a=0$ and using Leibniz’ rule
(the $\delta$-term is $a$-independent and drops) gives
\begin{align}
\partial_a^{k}S_a\Big|_{a=0}
=\sum_{n=0}^{k}\binom{k}{n}\,\Delta^{(n)}(0)\,
\partial_a^{\,k-n}\Big[\ln^{a-1}\!\frac{1}{x}\Big]_+\Big|_{a=0}.
\end{align}
Since $\Delta(0)=0$, the $n=0$ term vanishes. Moreover,
\begin{align}
\partial_a^{p}\,\ln^{a-1}\!\frac{1}{x}
=\ln^{a-1}\!\frac{1}{x}\,\Big(\ln\ln\frac{1}{x}\Big)^{p}
\quad\Rightarrow\quad
\partial_a^{p}\Big[\ln^{a-1}\!\frac{1}{x}\Big]_+\Big|_{a=0}
=\Big[\frac{\big(\ln\ln(1/x)\big)^{p}}{\ln(1/x)}\Big]_+,
\end{align}
and from $\Gamma(z+1)=z\Gamma(z)$ we have the recurrence
\begin{align}
\Delta^{(n)}(0)=n\,\Delta^{(n-1)}(1)\qquad(n\ge1),
\end{align}
with $\Delta(1)=1$ and $\Delta'(1)=\gamma_E$. Collecting these facts yields, for every $k\ge1$,
\begin{align}
\mathcal M^{-1}\!\left\{\ln^{k}\!\frac{1}{N}\right\}(x)
=\sum_{n=1}^{k}\binom{k}{n}\,n\,\Delta^{(n-1)}(1)\;
\Bigg[\frac{\big(\ln\ln\tfrac{1}{x}\big)^{\,k-n}}{\ln\tfrac{1}{x}}\Bigg]_+\,,
\end{align}
which can also be extended to the case $k=0$ by simply adding the delta term $\mathcal{M}^{-1}(1)=\delta(1-x)$. 

In the context of threshold resummation, we are mostly interested in the leading logarithms in $x$, so by performing an asymptotic expansion (near threshold $x\rightarrow 1^{-}$) of this expression, one gets for $p\geq 0$
\begin{align}
    \Bigg[\frac{\big(\ln\ln \tfrac{1}{x}\big)^{p}}{\ln \tfrac{1}{x}}\Bigg]_+
=
\Bigg[\frac{\ln^{p}(1-x)}{1-x}\Bigg]_+
+\mathcal{O}\!\Big([\ln^{p}(1-x)]\Big)
+\mathcal{O}\!\big((1-x)\,\ln^{p}(1-x)\big),
\end{align}
such that the leading power expansion can be written as
\begin{align}
    \mathcal M^{-1}\!\left\{\ln^{k}\!\frac{1}{N}\right\}(x)
\sim\sum_{n=1}^{k}\binom{k}{n}\,n\,\Delta^{(n-1)}(1)\;
\Bigg[\frac{\ln^{k-n}(1-x)}{1-x}\Bigg]_+\,,
\end{align}
so, for fixed $k$, we get the leading log contribution
\begin{align}
    \mathcal M^{-1}\!\left\{\ln^{k}\!\frac{1}{N}\right\}_{\textnormal{LL}}(x)=k\Bigg[\frac{\ln^{k-1}(1-x)}{1-x}\Bigg]_+\,.
\end{align}
Just keep in mind this is “LL within the $k$-block.” When one later organize a physical cross section order by order in $\alpha_s^n$, “LL at order $n$” refers to the highest $D_p(x):=\big[\ln^{p}(1-x)/(1-x)\big]_+$ that appears at that $\alpha_s^n$, which mixes contributions from several $k$-blocks, e.g.\ at $n=1$, LL is $D_1$, which comes from the $k=2$ block, while the $k=1$ block gives the NLL $D_0$.
\subsection{Convergence properties of the resummed cross section}
\label{sec:convergence_properties_Mellin}
The hadronic cross section can schematically be written as an integral over the threshold variable
\begin{align}\label{eq:hadronic_xsec_zintegral}
    \sigma(\tau,Q^2)=\int_{0}^{1}dx_1 \int_{0}^{1}dx_2 \int_{0}^{z_{\textnormal{max}}}dz\,\delta(zx_1x_2-\tau) f_1(x_1)f_2(x_2)w(z,Q^2)\,,
\end{align}
where we for later use have limited the $z$-integral by $z_{\textnormal{max}}$ even though the physical region is $z$-values in the interval $[0,1]$. Further, the sum over parton distributions and overall factors has been omitted as they are immaterial to the following discussion. After applying the delta function in the $z$-variable this can be written
\begin{align}\label{eq:factorizedhadronic}
    \sigma(\tau,Q^2)=\int_{\tau/z_{\textnormal{max}}}^1\frac{dx_1}{x_1}\int_{\tau/z_{\textnormal{max}}x_1}^1\frac{dx_2}{x_2}\,f_1(x_1)f_2(x_2)w\Big(z=\frac{\tau}{x_1x_2},Q^2\Big)\,,
\end{align}
which corresponds to our (\ref{eq:cs_PDFintegral}), and turns into  (\ref{eq:cs_PDFintegral_tilde}) when the $z$-variable is unbounded (see the discussion at the end of this section).

Since the resummed contribution to the cross section is obtained in Mellin space, we have to perform the inverse Mellin transform. The initial Mellin transform of \eqref{eq:hadronic_xsec_zintegral} using \eqref{eq:mellinoneside} is given by
\begin{align}
    \sigma(N)=F_{1}(N)F_{2}(N)W(N)\,,
\end{align}
where all the real space variables $x_1$, $x_2$ and $z$, are supposed to be defined on the interval $[0,1]$. For the resummed cross section we have
\begin{align}
    \sigma^{\textnormal{res}}(N)=F_{1}(N)F_{2}(N)W^{\textnormal{res}}(N)\,,
\end{align}
and the inverse is given by
\begin{align}
    \sigma^{\textnormal{res}}(\tau)&=\frac{1}{2\pi i}\int_{c-i\infty}^{c+i\infty}dN\,\tau^{-N}F_{1}(N)F_{2}(N)W^{\textnormal{res}}(N).
\end{align}
In this work we do not consider any Mellin space parametrization of the PDFs and therefore the inversion amounts to calculating\footnote{However, we do need to rewrite the PDF integrals to obtain numerical stability, see Section~\ref{sec:inverse_Mellin}.}
\begin{align}
    w^{\textnormal{res}}\big(z,Q^2\big)=\frac{1}{2\pi i}\int_{c-i\infty}^{c+i\infty}dN\,z^{-N}W^{\textnormal{res}}(N).
\end{align}

The full resummed coefficient function is given in Appendix~\ref{sec:expressions}, but for the following discussion we can without loss of generality consider the leading logarithmic contribution\footnote{The subleading contributions do not change the analytical properties of the resummed coefficient function, i.e.\ the branch points occur at the same place as for the leading logarithmic contribution.}
\begin{align}
    W^{\textnormal{res}}(N)=\exp\big(\ln{N}g_1\left(\lambda\right)\big)\,,
\end{align}
where
\begin{align}
    g_1\left(\lambda\right) &= \frac{A_1}{\pi b_0\lambda}\left(2\lambda+\left(1-2\lambda\right)\ln{\left(1-2\lambda\right)}\right)\,,\hspace{0.4cm}\lambda=b_{0}\alpha_s \ln N\,.
\end{align}
It is not difficult to show that $W^{\textnormal{res}}(N)$ is analytic for all complex $N$, except along two branch cuts: working on the principal branches of the logarithms, the factor $\ln N$ induces a cut along $(-\infty,0]$, while $\ln(1-2b_0\alpha_s \ln N)$ induces a cut along $[N_L,\infty)$, where
\begin{align}
    N_{L}=\exp\Big(\frac{1}{2b_0 \alpha_s}\Big)\,,
\end{align}
is the so-called Landau singularity. As a result, the inverse Mellin transform of the resummed coefficient function is not well-defined in the usual sense. That is, from the analysis in Section~\ref{eq:analytical_properties_Mellin}, the Mellin transform of a function $f(x)$ such that $|f(x)|\leq K x^{-\alpha}$ for all $x$, where $K$ and $\alpha$ are constants, is an analytic function of the complex variable $N$ in the half-plane $\textnormal{Re}(N) > \alpha$. Therefore, since $W^{\textnormal{res}}(N)$ has a branch cuts in the left \emph{and} right half-plane it cannot be the Mellin transform of any $w^{\textnormal{res}}(z)$ with $0\leq z\leq 1$. Further, the appearance of branch points implies that $W^{\textnormal{res}}(N)$ cannot be the Mellin transform of any regular function. The issue is two-fold: the series expansion in $\alpha_s$ does not converge and since each term in the series expansion is a distribution (some singular) in $x$-space, their (truncated) sum cannot straightforwardly yield a regular function. Resummation in Mellin space is based on formal manipulations and tacitly avoids addressing these issues, but they reappear as soon as one wants an $x$-space resummation formula. Thus, attempting to naively apply the inverse Mellin transform under the assumption that the integrand behaves well will inevitably fail.

To see the extent of the problem, we can write the exponent of the resummed coefficient function as a power expansion in $\alpha_s$. First, we observe that 
\begin{align}
    \ln(1-2b_0 \alpha_s \ln N)=\sum_{k=1}^{\infty}\frac{(-1)^{k+1}}{k}\Big(2b_0 \alpha_s \ln \frac{1}{N}\Big)^k\,,
\end{align}
such that we can write the resummed exponent $G(N):=\ln N g_1(\lambda)$ as the power series
\begin{align}
    G(N)=\sum_{k=1}^{\infty}c_k\alpha_{s}^{k} \ln^{k+1} \frac{1}{N} \,,
    \label{eq:lnWresN_series}
\end{align}
for some finite coefficients $c_k$ ($N$-independent). This series converges absolutely iff
\begin{align}
    \Big|2b_0 \alpha_s \ln \frac{1}{N}\Big|< 1\,
    .
\end{align}
However, the inverse integral involves values of $N$ outside this range, implying that the Mellin inverse of the series does not exist in the usual sense. That is, the Bromwich contour $N=c+iy$, $y\in(-\infty,\infty)$, does not keep $\Big|2b_0 \alpha_s \ln \frac{1}{N}\Big|< 1$. Thus, if we take the inverse of the series in (\ref{eq:lnWresN_series})
\begin{align}
    \frac{1}{2\pi i}\int_{c-i\infty}^{c+i\infty}dN\,z^{-N}\sum_{k=1}^{\infty}c_k\alpha_{s}^{k} \ln^{k+1} \frac
    {1}{N} \,,
\end{align}
and just blindly swap integration and summation such that each term could be calculated as
\begin{align}
    \sum_{k=1}^{\infty}c_k\alpha_{s}^{k} \frac{1}{2\pi i}\int_{c-i\infty}^{c+i\infty}dN\,z^{-N}\ln^{k+1} \frac
    {1}{N}\,,
\end{align}
and while each term in the series---understood in the distributional sense\footnote{Exactly of the form calculated in the previous section.}---is a well-defined inverse Mellin transform, the series does not converge. This is anticipated since, to interchange the order of integration and summation, one needs either \emph{absolute integrability} (Tonelli/Fubini) or a single integrable majorant on the contour dominating the partial sums (dominated convergence for series); neither condition holds here, so the interchange is invalid. However, as well-known\footnote{See~\cite{Forte:2006mi} for a detailed discussion.}, if one truncates to any finite logarithmic order, the $x$-space inverse (with $x$ below the Landau region) is perfectly well defined and the resulting perturbative series is under control. It is only when all logarithmic orders are included that the series diverges.

Going back to the full resummed coefficient function, one may consider the power expansion
\begin{align}
    W_{n}^{\textnormal{res}}(N)=\sum_{k=0}^{n}a_k(\ln N)\alpha_{s}^{k}\,,
\end{align}
where $a_k(\ln N)$ are polynomials in $\ln N$, and if one tried to take the limit $n\rightarrow\infty$, and calculate the inverse Mellin transform of this series expansion,
\begin{align}\label{eq:truncated series}
    w^{\textnormal{res}}\big(z,Q^2\big)=\lim_{n\rightarrow\infty}\sum_{k=0}^{n}\alpha_{s}^{k}\frac{1}{2\pi i}\int_{c-i\infty}^{c+i\infty}dN\,z^{-N}a_{k}(\ln N)\,,
\end{align}
one would again find that the series does not converge, and the inverse Mellin transform of the resummed coefficient function is not well-defined without some restriction.

This fact should not come as a surprise if one is acquainted with the challenges of perturbative series in quantum field theory. In most cases of physical interest, the perturbative series we work with are at best asymptotic rather than convergent. Hence, when trying to extract physically meaningful results out of the resummed expression, one must address these limitations. The appearance of the Landau pole underscores these shortcomings, but we know that this is an unavoidable feature of perturbative QCD, which is badly behaved at very low scales.



Thus, instead of being overly ambitious, one can instead \emph{define} the contour of the inverse Mellin transform to avoid the Landau pole singularity. Specifically, the integration contour may be restricted to pass to the left of the Landau pole, ensuring that the integral remains well-defined in the strip $(0,N_L)$. By doing so, one effectively exclude the non-perturbative region where the perturbative expansion breaks down, while retaining the resummed contributions that are valid in the perturbative regime.

This strategy is the content of the well-known \emph{Minimal Prescription}~\cite{Catani:1996yz}\footnote{An alternative is $x$-space Borel resummation (with a cutoff Borel transform) which treats plus-distributions directly and avoids any Mellin-inversion prescription; see, e.g.,~\cite{Forte:2006mi}.
}: The inverse Mellin transform of the resummed coefficient function may instead be defined as
\begin{align}
    w_{\textnormal{MP}}^{\textnormal{res}}\big(z,Q^2\big)=\frac{1}{2\pi i}\int_{c_{\textnormal{MP}}-i\infty}^{c_{\textnormal{MP}}+i\infty}dN\,z^{-N}W^{\textnormal{res}}(N),
\end{align}
where $c_\textnormal{MP}$ has to be chosen in $0<c_{\textnormal{MP}}<N_L$, and with this choice, it was shown that
\begin{itemize}
    \item $w_{\textnormal{MP}}^{\textnormal{res}}\big(z,Q^2\big)$ is free of Landau singularities.
    \item The series expansion~\ref{eq:truncated series} is asymptotic to $w_{\textnormal{MP}}^{\textnormal{res}}\big(z,Q^2\big)$, i.e. the $k$th order truncation of~\ref{eq:truncated series} differs from $w_{\textnormal{MP}}^{\textnormal{res}}\big(z,Q^2\big)$ by $O(\alpha_{s}^{k+1})$.
\end{itemize}

On the other hand, we observe the following problem: the strip of definition $0<c_{\textnormal{MP}}<N_L$ indicates that $w^{\textnormal{res}}\big(z)$ receives contributions from the non-physical domain $z>1$. However, as is established in~\cite{Catani:1996yz} these contributions are suppressed, and therefore in fact negligible. We have confirmed this numerically also in our calculations. 

In addition, we note that for the resummed hadronic cross section, $N$-space PDFs could introduce additional poles in $\textnormal{Re}(N)>0$, placing additional constraints on the strip of definition. We do not consider any $N$-space formulation of the PDFs and therefore do not have any control of possible poles. Thus, in general, we do not know what the exact strip of definition is for the full resummed cross section. In the minimal prescription, this problem is usually avoided by requiring that $c\geq2$, motivated by the Regge behaviour of structure functions;
one cannot have any singularity in structure functions above the Pomeron singularity, slightly above $N=1$.


\acknowledgments

We would like to thank Christopher Chang for discussions on numerical implementations, and collaboration on a closely related project to create a cross section calculation tool, as well as our other colleagues in the PLUMBIN' project for many interesting physics discussions.

This work was financed by the Research Council of Norway's grant number 323985, {\it PLUMBIN': Developing solvents for unclogging the calculational bottleneck in high-energy physics}. The numerical calculations were performed on resources provided by Sigma2 -- the National Infrastructure for High-Performance Computing and Data Storage in Norway, through allocation NN9284k.

\vspace{8pt}
\paragraph{Data Availability Statement.} This article has associated data in a data repository. The datasets are available at \url{https://doi.org/10.5281/zenodo.17093804}.

\vspace{8pt}
\paragraph{Code Availability Statement.} This article has no associated code or the code will not be deposited.

\vspace{8pt}
\paragraph{Open Access.} This article is distributed under the terms of the Creative Commons Attribution License (\href{https://creativecommons.org/licenses/by/4.0/}{CC-BY4.0}), which permits any use, distribution and reproduction in any medium, provided the original author(s) and source are credited.
 



\bibliographystyle{jhep}
\bibliography{biblio}

\end{document}